\documentclass{article}

\usepackage{arxiv}

\usepackage[utf8]{inputenc} 
\usepackage[T1]{fontenc}    
\usepackage{hyperref}       
\usepackage{url}            
\usepackage{booktabs}       
\usepackage{amsfonts}       
\usepackage{nicefrac}       
\usepackage{microtype}      
\usepackage{lipsum}
\usepackage{graphicx}

\usepackage{cite}
\usepackage{amsmath,amssymb,amsfonts}
\usepackage{algorithm}
\usepackage{algpseudocode}
\usepackage{algorithm}
\usepackage{algpseudocode}
\usepackage{graphicx}
\usepackage{subcaption}
\usepackage{textcomp}
\graphicspath{ {./images/} }

\title{Trajectory Reconstruction through a Gaussian Adaptive Selective Outlier Rejecting Smoother}

\author{
Arslan Majal \\
  Department Of Mechanical Engineering\\
  University of Wisconsin Madison\\
  \texttt{majal@wisc.edu} \\
   \And
 Aamir Hussain Chughtai \\
  Department Of Electrical Engineering\\
  Lahore University of Management Sciences\\
  \texttt{chughtaiah@gmail.com} \\
}

\begin{document}
\maketitle
\begin{abstract}
Trajectory Reconstruction (TR) is vital for accurately mapping movement patterns and validating analyses, especially in fields like robotics, biomechanics, and environmental tracking, where data might be missing or affected by outliers.
Improving Trajectory estimation by employing Gaussian smoothing techniques in the presence of non-Gaussian noise is the subject of this work. We consider the case where data is collected from independent sensors. A variational Bayesian (VB) based Unscented Raunch-Tung-Striebel smoothing (URTSS) scheme is proposed which adopts a vectorized weighing mechanism for the measurement covariance matrix to selectively remove contaminated measurements at each time step. To improve our outlier mitigation, we model our outlier characteristics as a Gamma distribution and dynamically learn the parameters of this distribution from data. We verify the performance of our proposed smoother by a range of simulations and experimental data. We also propose a robustness criterion for smoothers based on the Kullback-Leibler (KL) divergence and show that our proposed method complies with this criterion.

\end{abstract}

\section{Introduction}

TR involves estimating the complete path of a target over time using both historical and real-time sensor data, along with the physical principles governing the target’s movement \cite{garcia2015model}. TR plays a critical role in various fields such as air traffic management i.e. for performance assessment of different elements in surveillance systems and validating tracking tools \cite{besada2013atc} , ship route analysis for  enhancing collision avoidance measures \cite{liang2024aisclean} , human motion reconstruction \cite{ambrosio2001spatial}, accurate reconstruction of the path of a satellite \cite{takahashi2022trajectory} and marine exploration \cite{garcia2015model}, \cite{bai2020novel}. The accuracy of TR heavily depends on the quality of the collected data. However, challenges such as noisy data and missing data \cite{guo2021improved}, \cite{huang2023distributed}, \cite{huang2021event} often compromise the precision of the reconstruction.

In this work we are interested in treating the TR problem as an offline state estimation problem where the data and model information is already available. We further rely on the Rauch-Tung-Striebel (RTS) smoother framework which employs forward filtering and backward smoothing techniques to estimate the state at each time step, conditioned on all available measurements \cite{rauch1965maximum}. The (RTS) smoother is optimal for linear systems when both process and measurement noise follow Gaussian distributions and has sub-optimal extensions for non-linear systems as well \cite{sarkka2008unscented}. However, due to the use of low-cost sensors and complex environmental conditions, sensor noise often deviates from a Gaussian distribution \cite{plataniotis1997nonlinear}. Non-Gaussian noise contamination significantly diminishes the accuracy of Trajectory estimates. An established research direction is to develop methods that improve the performance of smoothers in non-Gaussian noise environments \cite{aravkin2017generalized}, while also enhancing the overall accuracy of TR. 

The RTS smoother is based on the least squares error function \cite{rauch1965maximum} which is very sensitive to outliers in the measurement vector. A variety of techniques utilizing a more resilient cost function \cite{aravkin2017generalized} are commonly employed to develop a more robust algorithm. M-estimation \cite{7855662} provides an important framework in this regard as it is able to cast the estimation problem as a linear and further non-linear regression problem \cite{karlgaard2015nonlinear}
. In Huber cost function \cite{muthukrishnan2010m} based M-estimation \cite{karlgaard2015nonlinear},\cite{8074807},\cite{5371933}, \cite{chang2012huber}, \cite{chang2015huber}  the traditional RTS smoother's quadratic cost function is modified by replacing it with a robust Huber objective function. This approach reduces the influence of outliers by down-weighting only the contaminated measurements, while treating the rest like a standard least-squares filter. By adjusting the weights instead of discarding the outliers completely, this method achieves a balance between robustness and maintaining high statistical efficiency. However, the performance of this method deteriorates when faced with high-intensity outliers/contaminants in the measurement vector. As pointed out in \cite{7814285} the influence function in Huber function based estimation does not redescend which may result in worse performance. Another important cost function comes from information learning methods as the Correntropy \cite{liu2007correntropy} \cite{santamaria2006generalized}. Correntropy is a similarity measurement between two random valuables and can be used to derive a robust trajectory estimation method through the Maximum Correntropy Criterion (MCC) \cite{wang2020maximum}, \cite{chen2017maximum} ,\cite{wang2016robust}. A major difficulty in using the MCC based methods is that Correntropy is highly dependent on the Kernel Bandwidth (KB) which can result in performance degradation\cite{chen2018mixture}. Another set of popular techniques for robust estimation employ Monte Carlo (MC) sampling \cite{doucet2001introduction}, such as Particle Filtering \cite{djuric2003particle},\cite{bilik2010mmse} to approximate the posterior state distribution by generating random samples, each with an associated weight. The computational strain associated with these sampling methods is an outstanding drawback for these techniques\cite{agamennoni2011outlier}.

A different commonly pursued approach for handling non-Gaussian noise in the measurement vector is to model the non-Gaussian noise \cite{huang2017novel}, \cite{6654132}, \cite{10155202} and incorporate these models into a robust estimation framework. This allows the system to better account for outliers and non-standard noise characteristics. An instance of this approach is the use of Gaussian Mixture Models (GMMs)\cite{he2024gaussian}, which can approximate any arbitrary distribution\cite{fan2022background}. In this approach, the measurement noise is decomposed into multiple Gaussian components using GMMs, allowing for the modeling of non-Gaussian distributions. A weighted combination of these models is then computed using the Interacting Multiple Model (IMM) framework\cite{blom1988interacting}. However, this approach is challenged by high computational complexity and the difficulty in determining the appropriate number of models needed to represent the unknown non-Gaussian measurement noise. Another prominent example of this strategy is
modeling the non-Gaussian measurement noise using a heavy tailed distribution and then dynamically estimating the parameters of this heavy tailed distribution. A prevalent trend is the use of VB approximations to enable tractable joint estimation of relevant parameters \cite{tzikas2008variational} for these distributions, such as the Student's t-distribution \cite{piche2012recursive}, \cite{huang2019novel}, \cite{wang2021novel}.

However, a major limitation of these methods is the use of a scalar weighting parameter for the measurement covariance matrix, which results in the rejection of the entire measurement vector whenever an outlier is detected. This can lead to the loss of valuable information. To address this issue, we introduced the Selective Outlier Rejecting Unscented RTS Smoother (SOR-URTSS) in \cite{10680401}, which employs a vectorized weighting mechanism. This approach allows selective rejection of individual components from the measurement vector, preserving relevant information.

Building on this, our current work presents an enhancement of the SOR-URTSS based on \cite{10430181}. Our proposed method adapts to outliers in the measurement vector by modeling the magnitude of the measurement covariance controlling weight as a Gamma distribution rather than just using a Bernoulli distribution for detecting outliers as in \cite{10680401}. The Gamma distribution rate parameter is estimated jointly and dynamically during the smoothing process from the measurement data. The use of a Gamma distribution to model our weight magnitude at each time step allows for a more precise suppression of outliers in the measurement vector. This results in more effective outlier mitigation and improved TR.

The key contributions of our approach are as follows:
\begin{itemize} 
\item We introduce a novel TR smoother for nonlinear systems with independent sensors, building on the State Space Model (SSM) from \cite{10680401}. Our adaptive approach enhances the robustness of TR by dynamically learning the parameters of outliers in the measurement vectors, offering improved performance over similar methods. 
\item We implement our proposed method using the Serial-Sigma Point Kalman Filter (S-SPKF), which enables it to achieve linear time complexity with respect to the number of measurement sensors, making it computationally efficient. 
\item We provide simulations and results using real-world data, comparing our method to other similar smoothers, and demonstrate its performance across a range of scenarios. 
\item We establish a robustness criterion for smoothers based on the Posterior Influence Function (PIF) \cite{duran2024outlier}  and demonstrate the robustness of our proposed method using this criterion. 
\end{itemize}

\section{Modeling Details}
We consider a nonlinear discrete-time SSM of a dynamic physical process described by
\begin{align}
\label{x_m}
&\mathbf{x}_k=\mathbf{f}(\mathbf{x}_{k-1})+\mathbf{q}_{k-1}\\
\label{y_m}
&\mathbf{y}_k=\mathbf{h}(\mathbf{x}_k)+\mathbf{r}_k
\end{align}
The subscript \(k\) denotes the time index, with \(\mathbf{x}_k \in \mathbb{R}^n\) as the state vector and \(\mathbf{y}_k \in \mathbb{R}^m\) as the measurement vector. The non-linear functions \(\mathbf{f}(.): \mathbb{R}^n \to \mathbb{R}^n\) and \(\mathbf{h}(.): \mathbb{R}^n \to \mathbb{R}^m\) define the process dynamics and observation models, respectively. The process noise \(\mathbf{q}_k \in \mathbb{R}^n\) and measurement noise \(\mathbf{r}_k \in \mathbb{R}^m\) are independent, white, and normally distributed with zero mean and covariances \(\mathbf{Q}_k\) and \(\mathbf{R}_k\), respectively. In this work, \(a_i\) denotes the \(i\)-th element of vector \(\mathbf{a}\), \(A^{ii}\) is the \(i\)-th diagonal element of matrix \(\mathbf{A}\), and \(\text{diag}(\mathbf{a})\) is a diagonal matrix with elements of \(\mathbf{a}\).
\\
Sensor observations can be corrupted by outliers, causing conventional filtering based on models \eqref{x_m}-\eqref{y_m} to fail. We assume observations come from independent sensors, so outliers are modeled independently for each dimension. To mitigate outlier effects on state estimation, we introduce an indicator vector \(\mathbf{\mathcal{I}}_k \in \mathbb{R}^m\) with Bernoulli elements. Specifically, \(\mathcal{I}_k^i\) can take values close to zero (\(\mathcal{I}_k^i = \epsilon\)) or one, where \(\mathcal{I}_k^i = \epsilon\) indicates an outlier in the corresponding dimension at time \(k\) and \(\mathcal{I}_k^i = 1\)  indicates the absence of an outlier. Outliers occur independently across dimensions and time, and \(\mathbf{\mathcal{I}}_k\) and \(\mathbf{x}_k\) are considered independent. Denoting \(\theta_k^i\) as the probability of no outlier in the \(i\)-th observation, the distribution of \(\mathbf{I}_k\) is given by:

\begin{align}
\label{I}
&p(\mathcal{I}_k)=\prod_{i=1}^mp(\mathcal{I}_k^i)=\prod_{i=1}^m(1-\theta_k^i)\delta(\mathcal{I}_k^i-\epsilon)+\theta_k^i\delta(\mathcal{I}_k^i-1) 
\end{align}

The measurement likelihood, conditioned on the current state \(\mathbf{x}_k\) and the indicator \(\mathbf{\mathcal{I}}_k\), independent of past observations \(\mathbf{y}_{1:k-1}\), is proposed to follow a Gaussian distribution given as

\begin{align}
\label{YI}
&p(\mathbf{y}_k|\mathbf{x}_k,\mathcal{I}_k)\\&=\mathcal{N}\left(\mathbf{y}_k|\mathbf{h}(\mathbf{x}_k),\mathbf{\Sigma}_k^{-1}\right)
\\
&=\prod_{i=1}^m\frac1{\sqrt{2\pi\operatorname{R}_k^{ii}/\mathcal{I}_k^i}}\exp\{-\frac{\left(\operatorname{y}_k^i-\operatorname{h}^i\left(\mathbf{x}_k\right)\right)^2}{2\operatorname{R}_k^{ii}}\mathcal{I}_k^i\}
\end{align}

Where \(\boldsymbol{\Sigma}_k = \mathbf{R}_k^{-1} \text{diag}(\mathbf{I}_k)\). We assume statistical independence in the nominal noise across measurement dimensions, especially for independently deployed sensors. Thus, \(\mathbf{R}_k\) is diagonal, allowing the distribution to be expressed as a product of univariate Gaussian distributions.

Considering \eqref{I} and \eqref{YI}, the modified measurement model incorporating the effect of outliers can be expressed as
\begin{align}
    \mathbf{y}_k=\mathbf{h}(\mathbf{x}_k)+\mathbf{v}_k
\end{align}
where the modified measurement noise assumes a Gaussian
mixture model as $\boldsymbol{\nu}_k\sim\sum_{\mathcal{I}_k}\mathcal{N}(\boldsymbol{\nu}_k|\boldsymbol{0},\boldsymbol{\Sigma}_k^{-1})p(\mathcal{I}_k).$

\section{Adaptive State Estimation}
To derive our smoother that employs an adaptive value for $\mathcal{I}_k^i = \epsilon$ we assume the following hierarchical prior distribution on $\mathcal{I}_k^i$ 
\[p(\mathcal{I}_k^i|b_k)=(1-\theta_i)\underbrace{f(a,b_k)(\mathcal{I}_k^i)^{a-1}e^{-b_k\mathcal{I}_k^i}}_{\mathcal{G}(\mathcal{I}_k^i|a,b_k)}+\theta_i\delta(\mathcal{I}_k^i-1) \]
where $\mathcal{I}_k^i$ follows a Gamma distribution with parameters $a$ and $b$ when an outlier occurs and we adaptively learn the parameter $b$ which captures the the effect of outliers in each measurement. The prior distribution for $b$ is also a Gamma distribution \cite{article} with parameters $A$ and $B$
\[p(b_k)=f(A,B)b_k^{A-1}e^{-Bb_k}\]
In order to derive our smoother we need to evaluate the joint conditional posterior distribution $p(\mathbf{x}_{1:T},\mathcal{I}_{1:T}, b_{1:T}|\mathbf{y}_{1:T})$. We employ the VB approximation which gives us the following marginal distributions as 
\[p(\mathbf{x}_{1:T},\mathcal{I}_{1:T}, b_{1:T}|\mathbf{y}_{1:T})\approx q(\mathbf{x}_{1:T})q(\mathcal{I}_{1:T})q( b_{1:T})\]
\begin{align}
\label{qx}
    &q(\mathbf{x}_{1:T})\propto\exp\left\{\left\langle\ln([p(\mathbf{x}_{1:T},\mathcal{I}_{1:T}, b_{1:T}|\mathbf{y}_{1:T}))\right\rangle_{q(\mathcal{I}_{1:T})q( b_{1:T})}\right\}\\
    \label{qi}
    &q(\mathcal{I}_{1:T})\propto\exp\left\{\left\langle\ln([p(\mathbf{x}_{1:T},\mathcal{I}_{1:T}, b_{1:T}|\mathbf{y}_{1:T}))\right\rangle_{q(\mathbf{x}_{1:T})q( b_{1:T})}\right\}\\
    \label{b}
    &\hat{b}_{1:T}=\underset{b_{1:T}}{\operatorname*{argmax}}\langle\ln(p(\mathbf{x}_{1:T},\mathcal{I}_{1:T}, b_{1:T}|\mathbf{y}_{1:T}))\rangle_{q(\mathcal{I}_{1:T})q(\mathbf{x}_{1:T})}
\end{align}
For the parameter $b$ we use the VB/EM theory with the assumption that $q(b_{1:T})= \prod_{k}^T \delta(b_k-\hat{b}_k)$.To evaluate $p(\mathbf{x}_{1:T},\mathcal{I}_{1:T}, b_{1:T}|\mathbf{y}_{1:T})$ we can use the Bayes theorem which gives us 
\begin{align}
    &p(\mathbf{x}_{1:T}, \mathcal{I}_{1:T}, b_{1:T}|\mathbf{y}_{1:T}) \propto \prod_{k}^T p(\mathbf{y}_k|\mathcal{I}_k, \mathbf{x}_k) p(\mathbf{x}_k|\mathbf{x}_{k-1}) p(\mathcal{I}_k|b_k) p(b_k)
\end{align}
Furthermore, we evaluate the log likelihood as
\begin{align}
&\ln(p(\mathbf{x}_{1:T}, \mathcal{I}_{1:T}, b_{1:T}|\mathbf{y}_{1:T})) \\
&= \sum_{k=1}^T \sum_{i=1}^m \left[ -0.5 \mathcal{I}_k^i (r_k^i(y_k^i, \mathbf{x}_k))^2 + 0.5 \ln(\mathcal{I}_k^i) \right. \\
&\quad + \ln\left( (1-\theta_i) f(a,b_k) (\mathcal{I}_k^i)^{a-1} e^{-b_k \mathcal{I}_k^i} + \theta_i \delta(\mathcal{I}_k^i-1) \right) \\
&\quad \left. -0.5 (r_0(\mathbf{y}_0, \mathbf{x}))^2 + (A-1)\ln(b_k) - B b_k + \text{constant} \right]
\end{align}
where $ r_k^i(y_k^i, \mathbf{x}_k)^2 = \|y_k^i - h^i(\mathbf{x}_k)\|_{R_k^{ii}}^{2}$ and ${R_k^{ii}}$ is the ith diagonal value of the measurement covariance matrix $\mathbf{R}_k$. We shall now be presenting the results of our derivation.
\subsection{Estimation of \texorpdfstring{$q(\mathbf{x}_{1:T})$}{q(x 1:T)}}
We can simplify \eqref{qx} as the following
\begin{align}
    q(\mathbf{x}_{1:T}) \propto \exp\left\{ \sum_{k = 1}^T \frac{-1}{2} (\mathbf{y}_k - \mathbf{h}(\mathbf{x}_k))^\top \mathbf{V}_k^{-1} (\mathbf{y}_k - \mathbf{h}(\mathbf{x}_k)) \right. \\
    \nonumber
    \left. - \frac{1}{2} (r_0(\mathbf{y}_0, \mathbf{x}))^2 \right\}
\end{align}
where $\mathbf{V}_k = \mathbf{R}_k \operatorname{diag}^{-1}\left(\langle \mathcal{I}_k \rangle_{q(\mathcal{I}_k)}\right)$. Due to VB conjugacy the posterior distribution $q(\mathcal{I}^i_k)$ has the same functional form as $p(\mathcal{I}_k^i|b_k)$ and thereby evaluated as 
\begin{align}
\label{Iq}
   \langle \mathcal{I}^i_k \rangle_{q(\mathcal{I}^i_k)} =  \Omega_k^i+(1-\Omega_k^i)\alpha/\beta_k^i~\forall~i>0
\end{align}
\subsection{Estimation of \texorpdfstring{$q(\mathcal{I}_{1:T})$}{q(I 1:T)}}

The equation \eqref{qi} can be evaluated to 
\begin{align}
\label{qi2}
&q(\mathcal{I}_{1:T}) \propto \prod_{k=1}^T \prod_{i=1}^m \left\{ (\mathcal{I}_k^i)^{0.5} e^{-0.5 W_k^{ii} \mathcal{I}_k^i} \right. \nonumber \\
    &\quad \left. \left[ (1-\theta_i) f(a, \hat{b}_k) (\mathcal{I}_k^i)^{a-1} e^{-\hat{b}_k \mathcal{I}_k^i} + \theta_i \delta(\mathcal{I}_k^i - 1) \right] \right\}
\end{align}
Where 
\begin{align}
\label{W}
W_{k}^{ii} = \frac{\langle y_k^i - h^i(\mathbf{x}_k) \rangle_{q(\mathcal{I}^i_k)}}{R_k^{ii}}
\end{align}
\eqref{qi2} can be further evaluated to get 
\begin{align}
&q\left(\mathcal{I}_{1:T}\right)= \prod_{k=1}^T \prod_{i=1}^m \overbrace{(1-\Omega_k^i) \underbrace{f(\alpha,\beta_k^i)(\mathcal{I}_k^i)^{\alpha-1} e^{-\beta_k^i \mathcal{I}_k^i}}_{q^1(\mathcal{I}_k^i)} + \Omega_k^i \delta(\mathcal{I}_k^i - 1)}^{q\left(\mathcal{I}_k^i\right)}
\end{align}
here $\alpha  = a + 0.5$, $\beta_k^i = 0.5W_k^{ii} + \hat{b}_k$ and 
\begin{align}
\label{Om}
\Omega_k^i=\frac{1}{1+\zeta\frac{\hat{b}_k^a}{(\beta_k^i)^\alpha}e^{0.5W_k^{ii}}}, \zeta=((1/\theta_i)-1)(\Gamma(\alpha)/\Gamma(a)) \forall i>0
\end{align}
\subsection{Estimation of \texorpdfstring{$\hat{b}_k$}{bk}}

We can maximize the right hand side of \eqref{b} by maximizing
\begin{align}
\label{qbb}
   \left\{ \sum_{k=1}^T \sum_{i=1}^m (1-\Omega_k^i) \left[\ln(f(a,b_k)) - b_k \langle \mathcal{I}_k^i \rangle_{q^1(\mathcal{I}_k^i)} \right] \right. \\
   \nonumber
   \left. + (A-1)\ln(b_k) - B b_k + \text{constant} \right\}
\end{align}
This can be achieved by maximizing \eqref{qbb} for each value of $k$ which gives
\begin{align}
&\bar{A}_k=A+\sum_{i=1}^ma(1-\Omega_k^i), \bar{B}_k=B+\sum_{i=1}^{m}(1-\Omega_k^i)\frac{\alpha}{\beta_k^i}
\\
\label{bk}
    &\hat{b}_k=\frac{\bar{A}_k-1}{\bar{B}_k}\quad\mathrm{s.t.}\quad\bar{A}_k>1
\end{align}
\section{Algorithm}
This section outlines the algorithmic steps for the S-SPKF implementation of our proposed algorithm, which consists of a forward and backward pass.
\\

\subsubsection{Forward Pass} This essentially constitutes the prediction and update steps of S-SPKF 
\subsubsection*{Prediction Step}

\begin{itemize} 
\item Generate the sigma points $\mathcal{X}_{k-1}^{(i)}$ as
\end{itemize} 
    \begin{align}
    \label{Sigma_Points_Start}
\mathcal{X}_{k-1}^{(0)}&=\mathbf{\hat{x}}_{k-1}\\
    \mathcal{X}_{k-1}^{(i)} & =\mathbf{\hat{x}}_{k-1}+\sqrt{n+\kappa}[\sqrt{\mathbf{\hat{P}}_{k-1}}]_i \quad i=1\ldots n\\
    \label{Sigma_Points_End}
    \mathcal{X}_{k-1}^{(i+n)} &=\mathbf{\hat{x}}_{k-1}-\sqrt{n+\kappa}[\sqrt{\mathbf{\hat{P}}_{k-1}}]_i\quad i=1\ldots n
    \end{align}
        where 
 $\mathbf{\hat{x}}_k$ and $\mathbf{\hat{P}}k$ denote the mean and covariance of the posterior distribution at time $k$, respectively. The superscript $(i)$ refers to the $i$th sigma point, $\kappa$ is the Unscented Transformation parameter \cite{julier1997new}, and $[\sqrt{\mathbf{\hat{P}}{k-1}}]i$ represents the $i$th column of the Cholesky decomposition of $\mathbf{\hat{P}}_{k-1}$.
        \begin{itemize}
        \item Pass the sigma points $\mathcal{X}_{k-1}^{(i)}$ through the dynamical function to get  $\mathcal{X}_{k}^{(i)-}$ and calculate the predicted state $\mathbf{\hat{x}}_k^-$ and state covariance $\mathbf{\hat{P}}_k^-$
    \end{itemize}
    \begin{align}
      \label{predicted_state}
    \mathcal{X}_{k}^{(i)-} &=\mathbf{f}\left(\mathcal{X}_{k-1}^{(i)}\right)\quad i=0\ldots2n ;  \quad \mathbf{\hat{x}}_k^-= \sum_{i=0}^{2n}\beta_i\mathcal{X}_{k}^{(i)-}\\
    \label{predicted_state_covariance}
    \mathbf{\hat{P}}_k^- &= \sum_{i=0}^{2n}\beta_i\left(\mathcal{X}_{k}^{(i)-} -\mathbf{\hat{x}}_k^-\right)\left(\mathcal{X}_{k}^{(i)-}-\mathbf{\hat{x}}_k^-\right)^T\\
    \label{eq:lk}
    \mathbf{L_k} &= \sum_{i=0}^{2n}\beta_i\left(\mathcal{X}_{k-1}^{(i)} -\mathbf{\hat{x}}_{k-1}\right)\left(\mathcal{X}_{k}^{(i)-}-\mathbf{\hat{x}}_{k}^-\right)^T
    \end{align}
\hspace{0.5cm}where $\beta_i\ = \frac\kappa{n+\kappa}$ for $i=0 $ or $\frac12\frac\kappa{n+\kappa}$ otherwise. $\mathbf{L_k}$ is used in the calculations of the subsequent Backward Pass 
\subsubsection*{Update Step}
    \begin{itemize}
    \item Pass the sigma points $\mathcal{X}_{k}^{(i)-}$ through the measurement model to obtain the predicted measurement sigma points $\mathcal{Y}{k}^{(i)}$, and compute the predicted measurement $\mathbf{v}_k$ \end{itemize}
   \begin{align}
   \mathcal{Y}_{k}^{(i)}&=\mathbf{h}\left(\mathcal{X}_{k}^{(i)-}\right)\quad i=0\ldots2n \quad ; \mathbf{v}_k = \sum_{i=0}^{2n}\beta_i\mathcal{Y}_{k}^{(i)}
   \end{align}
   
\begin{itemize}
    \item   For each measurement $l = 1 \hdots m$
\end{itemize}   
   \begin{align}
   & \mathrm{Y}_{k}\{l\} = \left[\ldots\sqrt{\beta_i}\left(\mathcal{Y}_{k}^{l(i)}-v^{l}_{k}\right)\ldots\right] \quad i=0\ldots2n
 \end{align}
 \begin{align}
   (\mathbf{C}_{k}\{l\})^{-1} &= (\mathbf{C}_k\{l-1\})^{-1}+\mathrm{Y}_{k}\{l\}^T\langle \mathbf{\mathcal{I}}^l_{k} \rangle_{q(\mathbf{\mathcal{I}}^l_{k})}(R_k^{ll})^{-1}\mathrm{Y}_{k}\{l\}\\
   \mathbf{d}_{k}\{l\} &= \mathbf{d}_{k-1}\{l-1\}+\mathrm{Y}_{k}\{l\}^T\langle \mathbf{\mathcal{I}}^l_{k} \rangle_{q(\mathbf{\mathcal{I}}^l_{k})}(R_k^{ll})^{-1}(y^l_{k}-v^{l}_{k})
   \end{align}

\noindent\hspace{0.5cm}
\parbox{\linewidth}{
    with $(\mathbf{C}_{k}\{0\})^{-1} =\textbf{1}, \mathbf{d}_{k}\{0\} =\textbf{0}$ 
    where $\mathcal{Y}_{k}^{l(i)}$ is the $l$th element of the $i$th sigma point at time $k$. The index $\{l\}$ refers to the vector or matrix in the $l$th cycle of the loop and $R_k^{ll}$ is the $l$th diagonal entry of $\mathbf{R}_k$.
}

   \begin{itemize}
       \item Compute the updated state and state covariance $\mathbf{\hat{x}}_{k} $ and $\mathbf{\hat{P}}_{k}$ as
   \end{itemize}
   \begin{align}
      &\mathcal{X}_k =\left[\ldots\sqrt{\beta_i}\left(\mathcal{X}_{k}^{(i)-}-\mathbf{\hat{x}}_k^-\right)\ldots\right] \quad i=0\ldots2n\\
      \label{filter_end}
&\mathbf{\hat{x}}_{k} =\mathbf{\hat{x}}_{k}^{-}+\mathcal{X}_k\mathbf{C}_{k}(m)\mathbf{d}_{k}(m),\quad\mathbf{\hat{P}}_{k}=\mathcal{X}_k\mathbf{C}_{k}(m)\mathcal{X}_k^{T}
\end{align}
\subsubsection{Backward Pass}
This involves the following computations by initializing the latest smoothing marginal parameters with $\mathbf{\hat{x}}_T^{s} = \mathbf{\hat{x}}_{T}$ and  $\mathbf{\hat{P}}_{T}^{s} = \mathbf{\hat{P}}_{T}$.  
\begin{itemize}
    \item For each value of $k$ compute the smoothing marginal density mean $\mathbf{\hat{x}}_k^{s}$ and covariance $\mathbf{\hat{P}}_{k}^{s}$ as 
\end{itemize}
\begin{align}
    \label{eq:gk}
    &\mathbf{G}_k = \mathbf{L}_k[\mathbf{\hat{P}}_{k+1}^{-}]^{-1} \\
        \label{eq:mus}
    &\mathbf{\hat{x}}_k^{s} = \mathbf{\hat{x}}_k + \mathbf{G}_k(\mathbf{\hat{x}}_{k+1}^{s} - \mathbf{\hat{x}}_{k+1}^{-}), \quad \mathbf{\hat{P}}_{k}^{s} = \mathbf{\hat{P}}_{k} + \mathbf{G}_k(\mathbf{\hat{P}}_{k+1}^{s} - \mathbf{\hat{P}}_{k+1}^{-})\mathbf{G}_{k}^{T}
    \end{align}

\begin{algorithm}
\caption{ASOR-URTSS}\label{alg:sors}
\textbf{Input}: $A,B,a, \mathbf{x}_0$, $\mathbf{y}_k$, $\mathbf{R}_k$, $\mathbf{Q}_k$ for $k = 1, \hdots, T$ \\
\textbf{Output}: $\mathbf{\hat{x}}^s_k$ for $k = 1, \hdots, T$
\begin{algorithmic} 
\State Initialize $\langle\mathcal{I}_k^i\rangle_{q(\mathcal{I}_k^i)} = 1$, $\theta^i = 0.5$, and $\hat{b}_k \ \forall \ i$ for $k = 1, \hdots, T$
\State Evaluate $\alpha = a + 0.5$ and $\zeta=(\frac{1}{\theta_{i}}-1)\frac{\Gamma(\alpha)}{\Gamma(a)}$
\While{\textit{not converged}}
\State Evaluate $\mathbf{\hat{x}}_k$ and $\mathbf{\hat{P}}_k$ for $k = 1, \hdots, T$ using  \eqref{Sigma_Points_Start} - \eqref{filter_end}
\State Evaluate $\mathbf{\hat{x}}_k^s$ and $\mathbf{\hat{P}}_k^s$ for $k = T, \hdots, 1$ using \eqref{eq:gk} - \eqref{eq:mus}
\State Evaluate $W_k^{ii}\ \forall i$ for $k = 1, \hdots, T$ using \eqref{W}
\State Evaluate $\beta_k^i = 0.5W_k^{ii} + \hat{b}_k \forall i$ for $k = 1, \hdots, T$
\State Evaluate $\Omega_k^i \forall i$ for $k = 1, \hdots, T$ using \eqref{Om}
\State Evaluate $\hat{b}_k$ for $k = 1, \hdots, T$ using \eqref{bk}
\State Evaluate $\langle \mathcal{I}^i_k \rangle_{q(\mathcal{I}^i_k)}$ $\forall i$  for $k = 1, \hdots, T$ using \eqref{Iq}
\EndWhile
\end{algorithmic}
\end{algorithm}

\section{PIFs for Smoothers}
In this section we present the PIFs \cite{duran2024outlier} for smoothers both in the Linear and Non-Linear domains. We define the PIF for a smoother within the time horizon $\{1,\hdots T\} $ and $k^c$ represent a time instance were the measurement $\mathbf{y}_k$ is corrupted, as following
\begin{align}
\mathrm{PIF}(\mathbf{y}_k^c,\mathbf{y}_{1:T \setminus \{k^c\}})=\mathrm{KL}\left(p(\mathbf{x}_k|\mathbf{y}_k^c,\mathbf{y}_{1:T \setminus \{k^c\}}\|p(\mathbf{x}_k|\mathbf{y}_{1:T})\right).
\end{align}

Where $\mathbf{y}_k^c$ is corrupted sensor measurement at time-step $k^c$. We further define a state estimation algorithm to be outlier robust if 
\[
\sup_{\mathbf{y}_k^c \in \mathbb{R}^m} |\operatorname{PIF}(\mathbf{y}_k^c, \mathbf{y}_{1:T \setminus \{k^c\}})| < \infty
\]
as $\|\mathbf{y}_k-\mathbf{y}_k^c\|_2\to \infty $

The standard formulation of the KL divergence between two Gaussian distributions with $n$-dimensional state vector is given as follows
\begin{align}
\nonumber
    &\mathrm{KL}(\mathcal{N}(\boldsymbol{\mu}_0, \boldsymbol{\Sigma}_0) \| \mathcal{N}(\boldsymbol{\mu}_1, \boldsymbol{\Sigma}_1)) = \frac{1}{2} \left( \mathrm{Tr}(\boldsymbol{\Sigma}_1^{-1} \boldsymbol{\Sigma}_0) - n \right. \\
    &\quad + (\boldsymbol{\mu}_1 - \boldsymbol{\mu}_0)^\top \boldsymbol{\Sigma}_1^{-1} (\boldsymbol{\mu}_1 - \boldsymbol{\mu}_0) + \ln\left(\frac{\det \boldsymbol{\Sigma}_1}{\det \boldsymbol{\Sigma}_0}\right) \big)
\end{align}
We first present the robustness criterion for Linear SSMs with weighted measurement covariance $\mathbf{R}_k$.

\subsection{Linear SSM with weighted measurement covariance \texorpdfstring{$\mathbf{R}_k$}{Rk}}

To robustify state estimation algorithms a weight $w_k$ is multiplied with the inverse of the measurement covariance $\mathbf{R}_t^{-1}$ to mitigate the effect of the outliers in measurement vector $\mathbf{y}_k^c$. The altered filter and smoother update equations are presented as
\begin{align}
&\hat{\mathbf{x}}_k=\mathbf{\hat{x}}_k^- +w_k\mathbf{K}_k\left(\mathbf{y}_k-\hat{\mathbf{y}}_k\right),\quad \hat{\mathbf{x}}_k^c=\mathbf{\hat{x}}_k^-+w_k^c\mathbf{K}_k^c\left(\mathbf{y}_k^c-\hat{\mathbf{y}}_k\right)\\
&\hat{\mathbf{P}}_k^{-1}=(\hat{\mathbf{P}}^-_{k})^{-1}+w_k\mathbf{H}_k^\top\mathbf{R}_k^{-1}\mathbf{H}_k, \quad (\hat{\mathbf{P}}_k^c)^{-1}=(\hat{\mathbf{P}}^-_k)^{-1} +w_k^c\mathbf{H}_k^\top\mathbf{R}_k^{-1}\mathbf{H}_k
\end{align}
\begin{align}
&\mathbf{\hat{x}}_k^{s} = \mathbf{\hat{x}}_k + \mathbf{G}_k(\mathbf{\hat{x}}_{k+1}^{s} - \mathbf{\hat{x}}_{k+1}^{-}), \quad \mathbf{\hat{x}}_k^{s,c} = \hat{\mathbf{x}}_k^c +\mathbf{G}^c_k(\mathbf{\hat{x}}_{k+1}^{s} -  \mathbf{\hat{x}}^{c-}_{k+1})\\
&\mathbf{\hat{P}}_{k}^{s} = \mathbf{\hat{P}}_{k} + \mathbf{G}_k(\mathbf{\hat{P}}_{k+1}^{s} - \mathbf{\hat{P}}_{k+1}^{-})\mathbf{G}_{k}^\top,\quad \mathbf{\hat{P}}_{k}^{s,c} = \mathbf{\hat{P}}_{k}^c + \mathbf{G}^c_t(\mathbf{\hat{P}}_{k+1}^{s} - \mathbf{\hat{P}}_{k+1}^{c-}){\mathbf{G}_t^c}^\top
\end{align}
where $\mathbf{\hat{x}}_k^{s,c}$ and $\mathbf{\hat{P}}_{k}^{s,c}$ represent the smoother update mean and covariance in the presence of an outlier in the sensor measurement and $\mathbf{H}_k$ represents the measurement model for $\mathbf{y}_k$.
Further the inverse of the smoother covariance is calculated as following

\begin{align}
    &(\mathbf{\hat{P}}_{k}^{s})^{-1} =  \mathbf{\hat{P}}_{k}^{-1} +  \mathbf{\hat{P}}_{k}^{-1}\mathbf{G}_k\mathbf{M}_k\mathbf{G}_k^\top\mathbf{\hat{P}}_{k}^{-1}\\
    &(\mathbf{\hat{P}}_{k}^{s,c})^{-1} =  (\mathbf{\hat{P}}_{k}^c)^{-1} +  (\mathbf{\hat{P}}_{k}^c)^{-1}\mathbf{G}_k\mathbf{M}_k\mathbf{G}_t^\top(\mathbf{\hat{P}}_{k}^c)^{-1}\\
\end{align}
where
\begin{align}
    &\mathbf{M}_k = (\mathbf{D}_k^{-1}  + \mathbf{G}_k^\top (\hat{\mathbf{P}}_k)^{-1} \mathbf{G}_k) ^{-1}\\
    &\mathbf{M}_k^c = ((\mathbf{D}_k^c)^{-1}  + {\mathbf{G}^c}_k^\top (\hat{\mathbf{P}}_k^c)^{-1} \mathbf{G}^c_k) ^{-1}\\
    &\mathbf{D}_k = \mathbf{\hat{P}}_{k+1}^{s} - \mathbf{\hat{P}}_{k+1}^{-}\\
    &\mathbf{D}_k^c = \mathbf{\hat{P}}_{k+1}^{s} - \mathbf{\hat{P}}_{k+1}^{c-}
\end{align}
\subsubsection{Linear SSM RTS smoother with weights such that \texorpdfstring{$0 < w_k < \infty$ , $w_k\|\mathbf{y}_k\|_2 < \infty$} and maintains the positive definiteness of \texorpdfstring{$\mathbf{\hat{P}}_{k}^{s,c}$} has a bounded PIF and is therefore outlier robust}
To show this we shall be adopting the results presented in \cite{duran2024outlier}.

\begin{align}
\nonumber
\label{pif}
&\operatorname{PIF}(\mathbf{y}_k^c, \mathbf{y}_{1:T \setminus \{k^c\}}) = \frac{1}{2} \Bigg( \underbrace{\text{Tr}\left((\mathbf{\hat{P}}_{k}^{s,c})^{-1}\mathbf{\hat{P}}_{k}^{s}\right) - n}_{(1)} \\
&+ \underbrace{( \mathbf{\hat{x}}_k^{s} -\mathbf{\hat{x}}_k^{s,c})^\mathsf{T} (\mathbf{\hat{P}}_{k}^{s,c})^{-1} ( \mathbf{\hat{x}}_k^{s} -\mathbf{\hat{x}}_k^{s,c})}_{(2)} + \underbrace{\text{ln}\left(\frac{\det \mathbf{\hat{P}}_{k}^{s,c}}{\det\mathbf{\hat{P}}_{k}^{s}}\right)}_{(3)} \Bigg)
\end{align}

\begin{align}
    &(1) = \text{Tr}\left((\mathbf{\hat{P}}_{k}^{s,c})^{-1}\mathbf{\hat{P}}_{k}^{s}\right) - n \leq \text{Tr}(\mathbf{\hat{P}}_{k}^{s,c})^{-1}\text{Tr}(\mathbf{\hat{P}}_{k}^{s}) -n
\end{align}
we employ the fact that for two positive definite matrices $\mathbf{A}$ and $\mathbf{B}$, $\text{Tr}\left(\mathbf{A} \mathbf{B}\right) \leq \text{Tr}(\mathbf{A})\text{Tr}(\mathbf{B})$. We can set $\text{Tr}(\mathbf{\hat{P}}_{k}^{s}) = C_1$ since it is not dependent on $\mathbf{y}_k^c$.
\begin{align}
    &\text{Tr}((\mathbf{\hat{P}}_{k}^{s,c})^{-1}) = \text{Tr}((\mathbf{\hat{P}}_{k}^c)^{-1} + (\mathbf{\hat{P}}_{k}^c)^{-1}\mathbf{G}^c_k\mathbf{M}_k^c{\mathbf{G}^c_k}^\top(\mathbf{\hat{P}}_{k}^c)^{-1})\\
    &\text{Tr}((\mathbf{\hat{P}}_{k}^{s,c})^{-1}) = \text{Tr}((\mathbf{\hat{P}}_{k}^c)^{-1}) + \text{Tr}((\mathbf{\hat{P}}_{k}^c)^{-1}\mathbf{G}^c_k\mathbf{M}_k^c{\mathbf{G}^c_k}^\top(\mathbf{\hat{P}}_{k}^c)^{-1}))
\end{align}
We can see here that $\text{Tr}((\mathbf{\hat{P}}_{k}^c)^{-1}) = \mathrm{Tr}\left( (\hat{\mathbf{P}}^-_k)^{-1}\right)+w_k^c \mathrm{Tr}\left(\mathbf{H}_k^\top\mathbf{R}_k^{-1}\mathbf{H}_k\right)$. For a constant value of $0 \leq w_k^c \leq C_2 < \infty$ we can set $\text{Tr}((\mathbf{\hat{P}}_{k}^c)^{-1}) = C_3$ \cite{duran2024outlier}.\\


\begin{align}
    &\text{Tr}((\mathbf{\hat{P}}_{k}^{s,c})^{-1}) \leq C_3 + \text{Tr}((\mathbf{\hat{P}}_{k}^c)^{-1}\mathbf{G}^c_k\mathbf{M}_k^c{\mathbf{G}^c_t}^\top(\mathbf{\hat{P}}_{k}^c)^{-1})  
\end{align}
we can further expand this as
\begin{align}\nonumber
    &= C_3 + \text{Tr}( \hat{(\mathbf{P}}^-_{k})^{-1}\mathbf{G}^c_k\mathbf{M}_k^c{\mathbf{G}^c_k}^\top (\hat{\mathbf{P}}^-_{k})^{-1})  + w_k^c \text{Tr}((\hat{\mathbf{P}}^-_{k})^{-1}\mathbf{G}^c_k\mathbf{M}_k^c{\mathbf{G}^c_k}^\top\mathbf{H}_k^\top\mathbf{R}_k^{-1}\mathbf{H}_k) + w_k^c \text{Tr}(\mathbf{H}_k^\top\mathbf{R}_k^{-1}\mathbf{H}_k\mathbf{G}^c_t\mathbf{M}_k^c{\mathbf{G}^c_k}^\top (\hat{\mathbf{P}}^-_{k})^{-1})\\
    & + (w_k^c)^2 \text{Tr}(\mathbf{H}_k^\top\mathbf{R}_k^{-1}\mathbf{H}_k\mathbf{G}^c_k\mathbf{M}_k^c{\mathbf{G}^c_k}^\top\mathbf{H}_k^\top\mathbf{R}_k^{-1}\mathbf{H}_k))
\end{align}
Due to the bounded nature of  $\mathbf{M}_k^c$ and $\mathbf{G}^c_k$, resulting from the boundedness of $\mathbf{\hat{P}}_{k+1}^{c-}$ in the forward pass when $0 \leq w_k^c \leq \infty$ \cite{duran2024outlier} we can conclude that the traces in the subsequent terms remains bounded.
\begin{align}
    & = C_3 + C_4 -n\\
    \nonumber
    & = C_5
\end{align}

\begin{align}
    (2) &= (\| \mathbf{\hat{x}}_k^{s} - \mathbf{\hat{x}}_k^{s,c} \|_{\mathbf{\hat{P}}_{k}^{s,c}})^2 \\
    &= (\| \hat{\mathbf{x}}_k - \hat{\mathbf{x}}_k^c + \mathbf{G}_k (\mathbf{\hat{x}}_{k+1}^{s} - \mathbf{\hat{x}}_{k+1}^{-})  -  \mathbf{G}^c_k (\mathbf{\hat{x}}_{k+1}^{c,s} - \mathbf{\hat{x}}_{k+1}^{c-})\|_{\mathbf{\hat{P}}_{k}^{s,c}})^2 \\
    &\leq \left(\| \hat{\mathbf{x}}_k - \hat{\mathbf{x}}_k^c\|_{\mathbf{\hat{P}}_{k}^{s,c}}     + \| \mathbf{G}_k (\mathbf{\hat{x}}_{k+1}^{s} - \mathbf{\hat{x}}_{k+1}^{-}) \|_{\mathbf{\hat{P}}_{k}^{s,c}} + \|\mathbf{G}^c_k (\mathbf{\hat{x}}_{k+1}^{c,s} - \mathbf{\hat{x}}_{k+1}^{c-})\|_{\mathbf{\hat{P}}_{k}^{s,c}} \right)^2
\end{align}
here it can  be shown that $\|\hat{\mathbf{x}}_k - \hat{\mathbf{x}}_k^c \|_{\mathbf{\hat{P}}_{k}^{s,c}} \leq \infty$ as in \cite{duran2024outlier} and further due to $\leq w_k^c \leq \infty$, $G_k^c$ and $\mathbf{\hat{x}}_{k+1}^{c-}$ also remain bounded in the forward pass due to the boundedness of forward pass estimate $\mathbf{x}^c_k$  from \cite{duran2024outlier}. In conclusion we can say that $(2) = C_6$
\begin{align}
    &(3) = \text{ln} \bigg(\frac{\det \mathbf{\hat{P}}_{k}^{s,c}}{\det\mathbf{\hat{P}}_{k}^{s}} \bigg) = \text{ln}\bigg( \frac{1}{\det\mathbf{\hat{P}}_{k}^{s}}\bigg) + 
    \text{ln}\bigg( \det(\mathbf{\hat{P}}_{k}^{s,c})\bigg)
\end{align}
Since $\text{ln}\bigg( \frac{1}{\det\mathbf{\hat{P}}_{k}^{s}}\bigg)$ does not depend on $\mathbf{y}_k^c$ we can set it equal to a constant $C_{7}$

Further we can see that $\mathbf{\hat{P}}_{k}^{s,c}$ can be written as
\begin{align}
&\mathbf{\hat{P}}_{k}^{s,c} = \mathbf{\hat{P}}_{k}^c + \mathbf{G}^c_t\mathbf{\hat{P}}_{k+1}^{s}{\mathbf{G}_t^c}^\top + (- \mathbf{G}^c_t\mathbf{\hat{P}}_{k+1}^{c-}{\mathbf{G}_t^c}^\top)
\end{align}
Here $\mathbf{\hat{P}}_{k}^c$ is positive definite \cite{duran2024outlier} and it can be shown that $\mathbf{G}^c_t\mathbf{\hat{P}}_{k+1}^{s}{\mathbf{G}_t^c}^\top$ is positive definite while $(- \mathbf{G}^c_t\mathbf{\hat{P}}_{k+1}^{c-}{\mathbf{G}_t^c}^\top)$ is negative definite, thereby making the nature of $\mathbf{\hat{P}}_{k}^{s,c}$ positive indefinite. To alleviate this, it must be ensured that $\mathbf{\hat{P}}_{k}^{s,c}$ is maintained as positive definite either through the robustness of the smoother or some external nearest positive definite approximating algorithm. This bounds the determinant, $\det (\mathbf{\hat{P}}_{k}^{s,c}) = C_8$.

\begin{align}
    &\text{PIF}(\mathbf{y}_t^c, \mathbf{y}_{1:T \setminus \{t^c\}}) = \frac{1}{2}\big( (1) +(2) + (3)\big) \\
    &\leq \frac{1}{2}\big(C_5 +  C_6 + C_7 + C_8 \big) \leq \infty
\end{align}

\subsection{Non-linear SSM with weighted measurement covariance \texorpdfstring{$\mathbf{R}_k$}{Rk}}

In a similar fashion as \cite{duran2024outlier}, we can easily extend this proof for smoother PIFs to nonlinear systems by employing the the Unscented Transform (UT) or the Extended Kalman Filter (EKF) \cite{sarkka2023bayesian} to evaluate individual entries in our calculations but the final results maintain their boundedness under the criterion established for linear SSMs.

\section{Robustness of ASOR-URTSS}


In this section, we evaluate the robustness of the ASOR-URTSS with respect to the PIF criterion established earlier, specifically: \( 0 < w_k < \infty \) , \( w_k \|\mathbf{y}_k\|_2 < \infty \) and positive definiteness of 
$\mathbf{\hat{P}}_{k}^{s,c}$
The ASOR-URTSS utilizes a vector of Bernoulli random variables, denoted as \( \mathbf{\mathcal{I}}_k \), to weight \( \mathbf{R}_k^{-1} \). Since each element of this vector is bounded, \( 0 \leq \mathcal{I}_k^i \leq 1 \), it satisfies the first criterion. Additionally, from equations \eqref{Iq} to \eqref{bk}, it can be observed that as the divergence between \( y_k^i \) and the predicted measurement \( h^i(\mathbf{x}_k) \) increases, \( W_k^{ii} \rightarrow \infty \) for all \( i \), resulting in \( \langle \mathbf{\mathcal{I}}^i_{k} \rangle_{q(\mathbf{\mathcal{I}}^i_{k})} \rightarrow 0 \) for all \( i \). This ensures that \( \langle \mathbf{\mathcal{I}}^i_{k} \rangle_{q(\mathbf{\mathcal{I}}^i_{k})} \|\mathbf{y}_k\|_2 < \infty \) for all \( i \), thus satisfying the second criterion. We also employ a nearest positive definiteness method to assert the positive definiteness of our $\mathbf{\hat{P}}_{k}^{s,c}$ at each time step.

\section{Simulations}
To demonstrate the performance improvements and key properties of our proposed Adaptive Selective Outlier Rejecting Unscented Rauch-Tung-Striebel Smoother (ASOR-URTSS), we conducted simulations across three distinct scenarios. We compared our method against the Selective Outlier Rejecting Unscented Rauch-Tung-Striebel Smoother (SOR-URTSS), as described in \cite{10680401}, and the Recursive Outlier Rejecting Unscented Rauch-Tung-Striebel Smoother (ROR-URTSS), which we derived based on the Recursive Outlier Rejecting (ROR) filter from \cite{piche2012recursive}. The ROR-URTSS mitigates outliers by applying a scalar weighting to the measurement covariance matrix. Additionally, to provide a clear perspective on the performance gains, we compared our results with an Ideal Unscented Rauch-Tung-Striebel Smoother (Ideal-URTSS), which assumes perfect knowledge of the outliers and removes them completely from the measurement vector. We also present results comparing the PIFs of SOR-URTSS and ASOR-URTSS for varying levels of divergence between corrupted and predicted measurements. 
\subsection{Setup}

The simulations were conducted on a Windows machine equipped with a 13th-generation Intel i9 processor and 32 GB of RAM. The generated data simulate a target tracking problem, with the state dynamics modeled as follows.
\begin{align}
    &\mathbf{x_k} = \begin{pmatrix}
        1& \frac{\sin{\omega_{k-1} \Delta t}}{\omega_{k-1}}& 0& \frac{\cos{\omega_{k-1} \Delta t} - 1}{\omega_{k-1}} &0 \\
        0& \cos{\omega_{k-1} \Delta t}& 0 & -\sin{\omega_{k-1} \Delta t}& 0
        \\
        0& \frac{1-\cos{\omega_{k-1} \Delta t}}{\omega_{k-1}}& 1& \frac{\sin{\omega_{k-1} \Delta t}}{\omega_{k-1}} & 0
        \\
        0& \sin{\omega_{k-1} \Delta t} &0&\cos{\omega_{k-1} \Delta t}&0
        \\
        0& 0& 0& 0& 0
    \end{pmatrix}\mathbf{x_{k-1}} + \mathbf{q_{k-1}}
\end{align}
where the state vector $\mathbf{x_k} = [a_k, \dot{a}_k, b_k, \dot{b}_k , \omega_k]$ is comprised of the 2D coordinates of the moving target $(a_k, b_k)$, the corresponding velocities $(\dot{a}_k, \dot{b}_k)$ and the angular velocity $\omega_k$ at the time instant $k$. Moreover, $\Delta t$ is the sampling period and $\mathbf{q}_{k-1} \sim \mathcal{N}(0, \mathbf{Q}_{k-1})$ where
\begin{align}
\nonumber
    &\mathbf{Q}_{k-1} = \begin{pmatrix}
        \eta_{1} \mathbf{M}& 0 & 0 
        \\
        0 & \eta_1 \mathbf{M} & 0
        \\
        0& 0& \eta_{2}
    \end{pmatrix}, \mathbf{M} = \begin{pmatrix}
        \frac{\Delta t^{3}}{3}& \frac{\Delta t^{2}}{2}
        \\
        \frac{\Delta t^{2}}{2} & \Delta t
    \end{pmatrix}
\end{align}

We assume the presence of $m/2$ bearing sensors and $m/2$ range sensors. The $j$th bearing sensor is located at coordinates $\left(a^{\theta_j} = 350(j-1), b^{\theta_j} = 350(j \mod 2)\right)$, while the $j$th range sensor is positioned at $\left(a^{\rho_j} = 350(j-1), b^{\rho_j} = 350((j+1) \mod 2)\right)$. The corrupted observations for both range and bearing are given as follows.
\begin{align}
     {y_k^{\rho_j}} = \sqrt{({a_k} - a^{\rho_j})^2 + ({b_k} - b^{\rho_j})^2} + r^{\rho_j}_{k} + \Lambda_k^{\rho_j} o^{\rho_j}_{k}\\
     {y_k^{\theta_j}} = \operatorname{atan2}({b_k} - b^{\theta_{j}} , {a_k} - a^{\theta_{j}}) + r^{\theta_{j}}_{k} + \Lambda_k^{\theta_j} o^{\theta_{j}}_{k}
\end{align}
 where $r^{\rho_j}_{k}$ and $r^{\theta_{j}}_{k}$ are the nominal noise in the $ j $th  range and angle sensors respectively. The outlier addition in each dimension is dictated using Bernoulli random variables $\Lambda_k^{\rho_j}$ and $\Lambda_k^{\theta_j}$ with probabilities of each being $1$  equal to $\lambda$ and normally distributed random variables $o^{\rho_j}_{k}$ and $o^{\theta_{j}}_{k}$.


For our simulations, we sample the initial state $\mathbf{x}_0$ from the distribution $\mathcal{N} \left([0, 10, 0, -5, -3, \frac{\pi}{180}]^{T}, \vartheta \mathbf{Q}_k \right)$, with the parameter $\vartheta$ set to 10. The simulation parameters are as follows: time step $\Delta t = 1$, $\eta_1 = 0.1$, $\eta_2 = 1.75 \times 10^{-4}$, and the initial covariance $\mathbf{P}_0 = \vartheta \mathbf{Q}_k$. The simulation runs for $T = 100$ time steps. The range noise $r^{\rho_j}_{k}$ is sampled from $\mathcal{N}(0, 10)$, and the bearing noise $r^{\theta_j}_{k}$ from $\mathcal{N} \left(0, \left(\frac{0.2\pi}{180}\right)^2 \right)$. The outlier terms are set as $o^{\rho_j}_{k} = \varsigma r^{\rho_j}_{k}$ and $o^{\theta_j}_{k} = \varsigma r^{\theta_j}_{k}$ where $\varsigma = \sqrt{1000}$.

For the unscented transform, the parameters are set as $\alpha = 1$, $\beta = 2$, $\kappa = 0$, and $\epsilon = 10^{-6}$. Benchmark methods are implemented using their originally reported parameters. Each scenario is evaluated using 50 Monte Carlo runs.
\subsection{Simulation 1}
\begin{figure*}[ht]
    \centering
    \begin{minipage}[t]{0.24\textwidth}
        \centering
        \includegraphics[width=\textwidth]{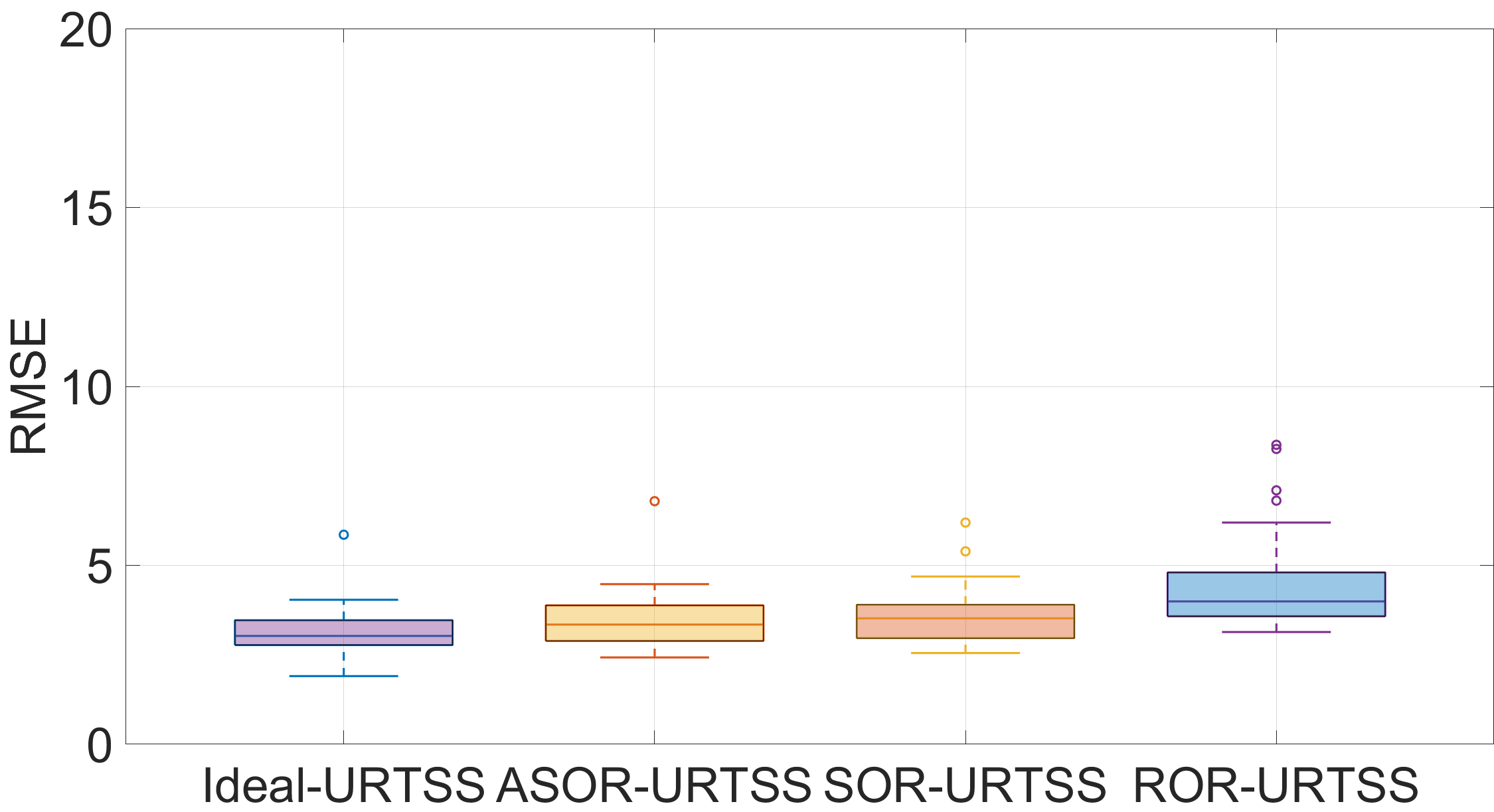}
        \subcaption{$\lambda = 0.2$}\label{fig:a1}
    \end{minipage}\hfill
    \begin{minipage}[t]{0.24\textwidth}
        \centering
        \includegraphics[width=\textwidth]{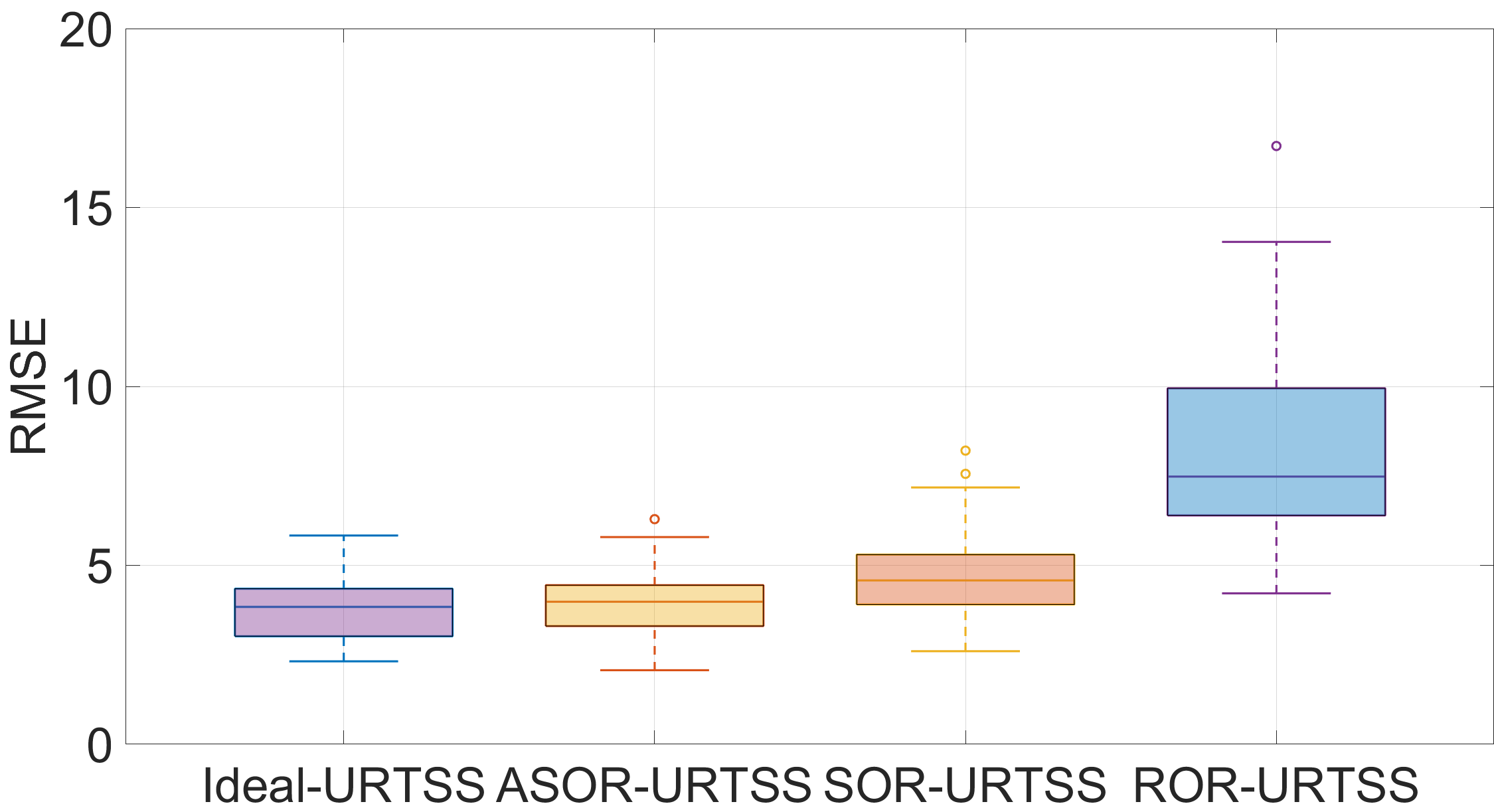}
        \subcaption{$\lambda = 0.4$}\label{fig:b1}
    \end{minipage}\hfill
    \begin{minipage}[t]{0.24\textwidth}
        \centering
        \includegraphics[width=\textwidth]{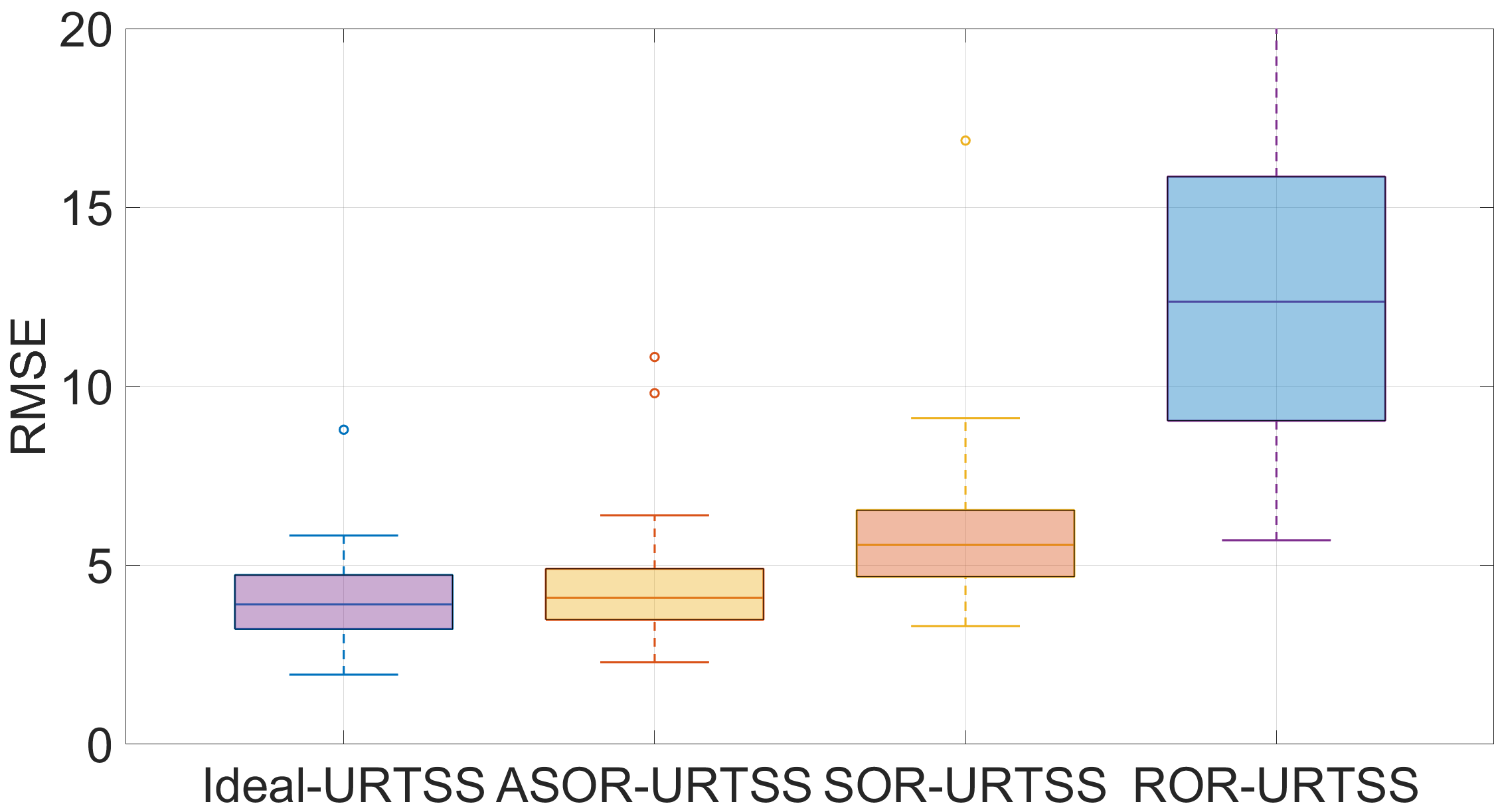}
        \subcaption{$\lambda = 0.5$}\label{fig:c}
    \end{minipage}\hfill
    \begin{minipage}[t]{0.24\textwidth}
        \centering
        \includegraphics[width=\textwidth]{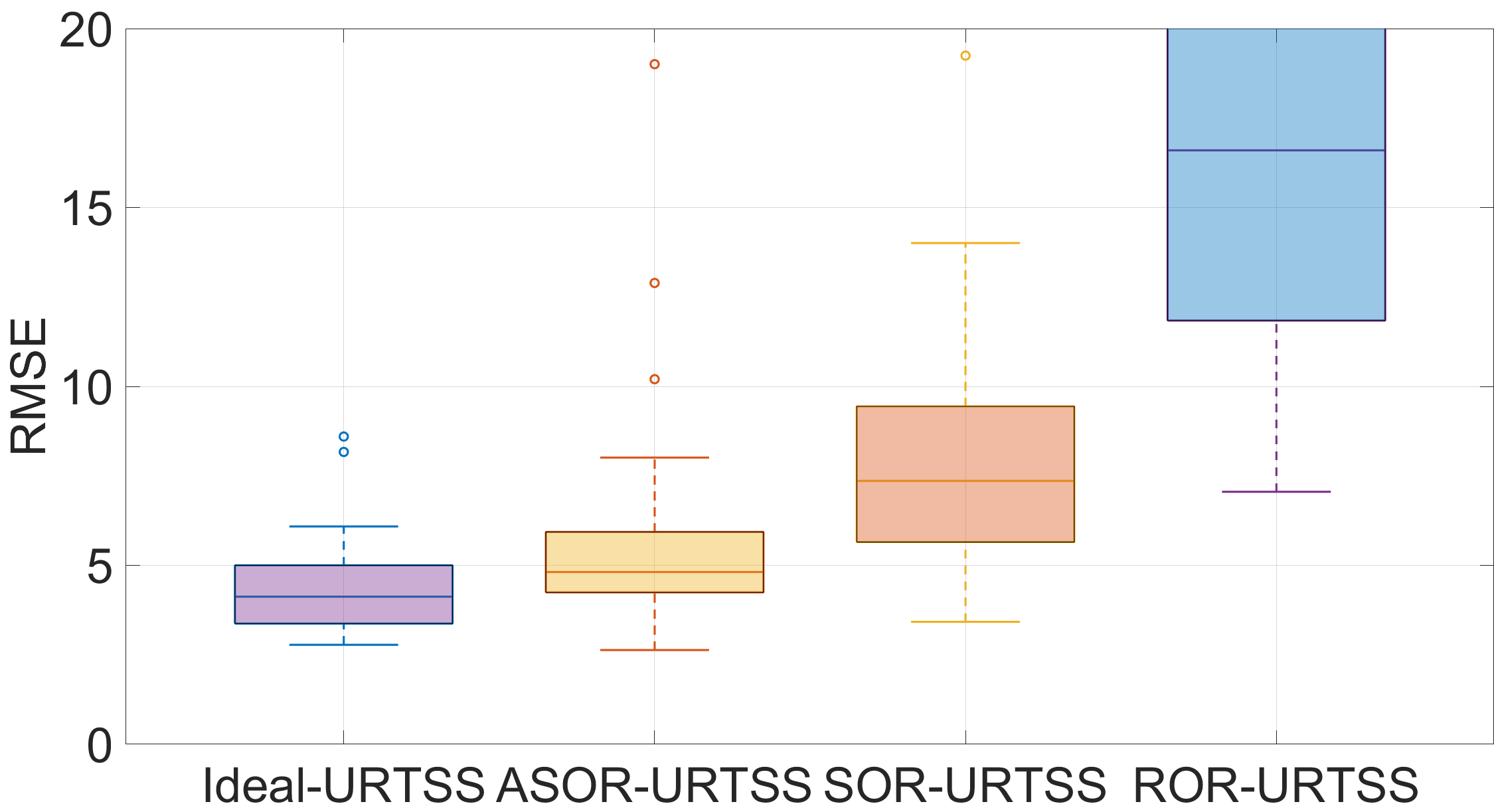}
        \subcaption{$\lambda = 0.6$}\label{fig:d}
    \end{minipage}
    
    \caption{Results for Simulation 1}
    \label{fig:s1}
\end{figure*}

The goal of this simulation is to showcase the improvement of our proposed smoother against increasing value of $\lambda$ i.e. $\lambda \in \{0.2, 0.4, 0.5, 0.6\}$. The results presented in Figure $\ref{fig:s1}$ highlight that the ASOR-URTSS outperforms all non-ideal smoothers, particularly at higher values of $\lambda$, This improvement can be attributed to the ability of the ASOR-URTSS to learn the characteristics of outliers by estimating $\hat{b}_k$ for each time step.

\subsection{Simulation 2}
\begin{figure*}[ht]
    \centering
    \begin{minipage}[t]{0.24\textwidth}
        \centering
        \includegraphics[width=\textwidth]{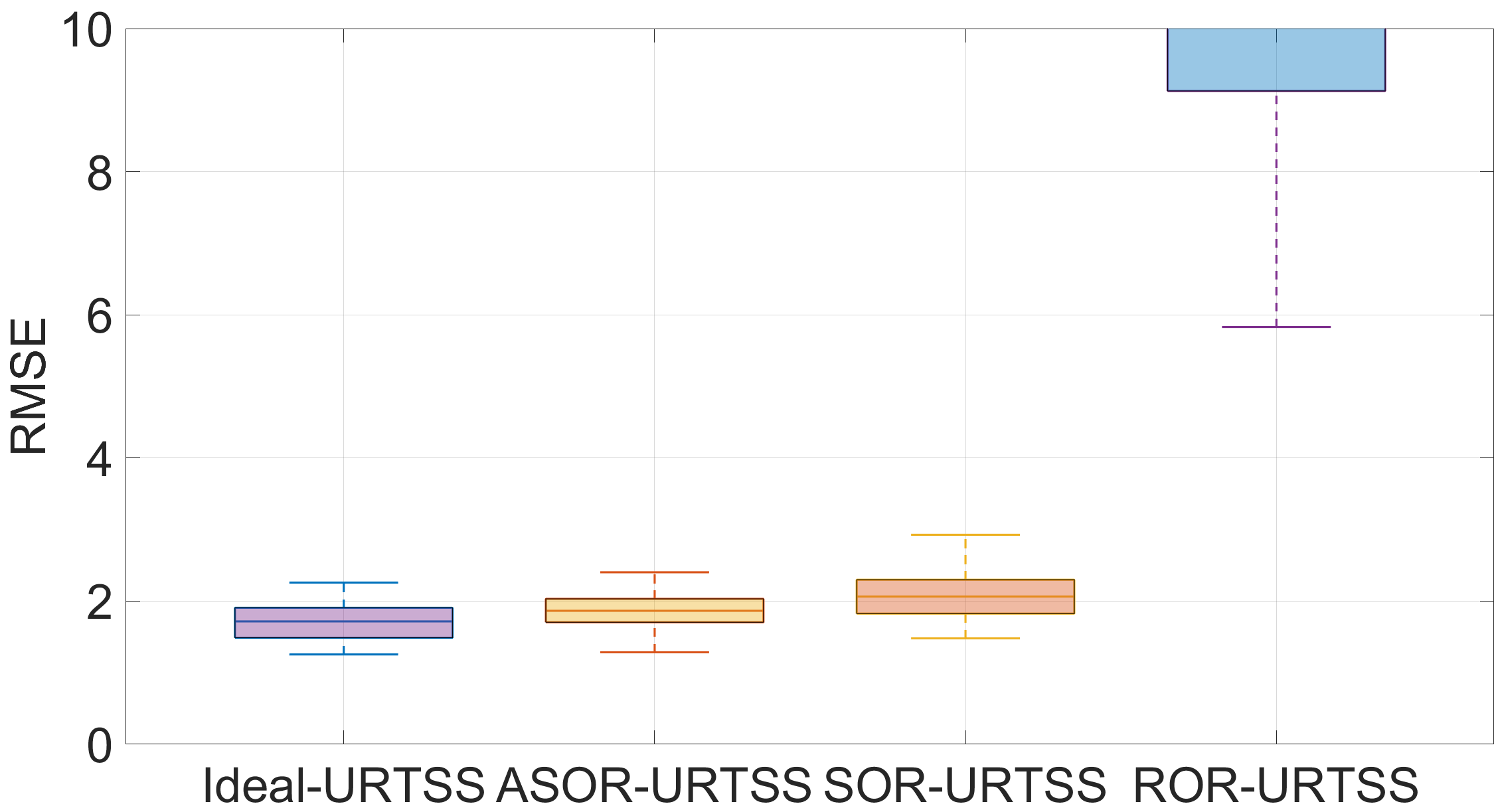}
        \subcaption{$m = 50$}\label{fig:a2}
    \end{minipage}\hfill
    \begin{minipage}[t]{0.24\textwidth}
        \centering
        \includegraphics[width=\textwidth]{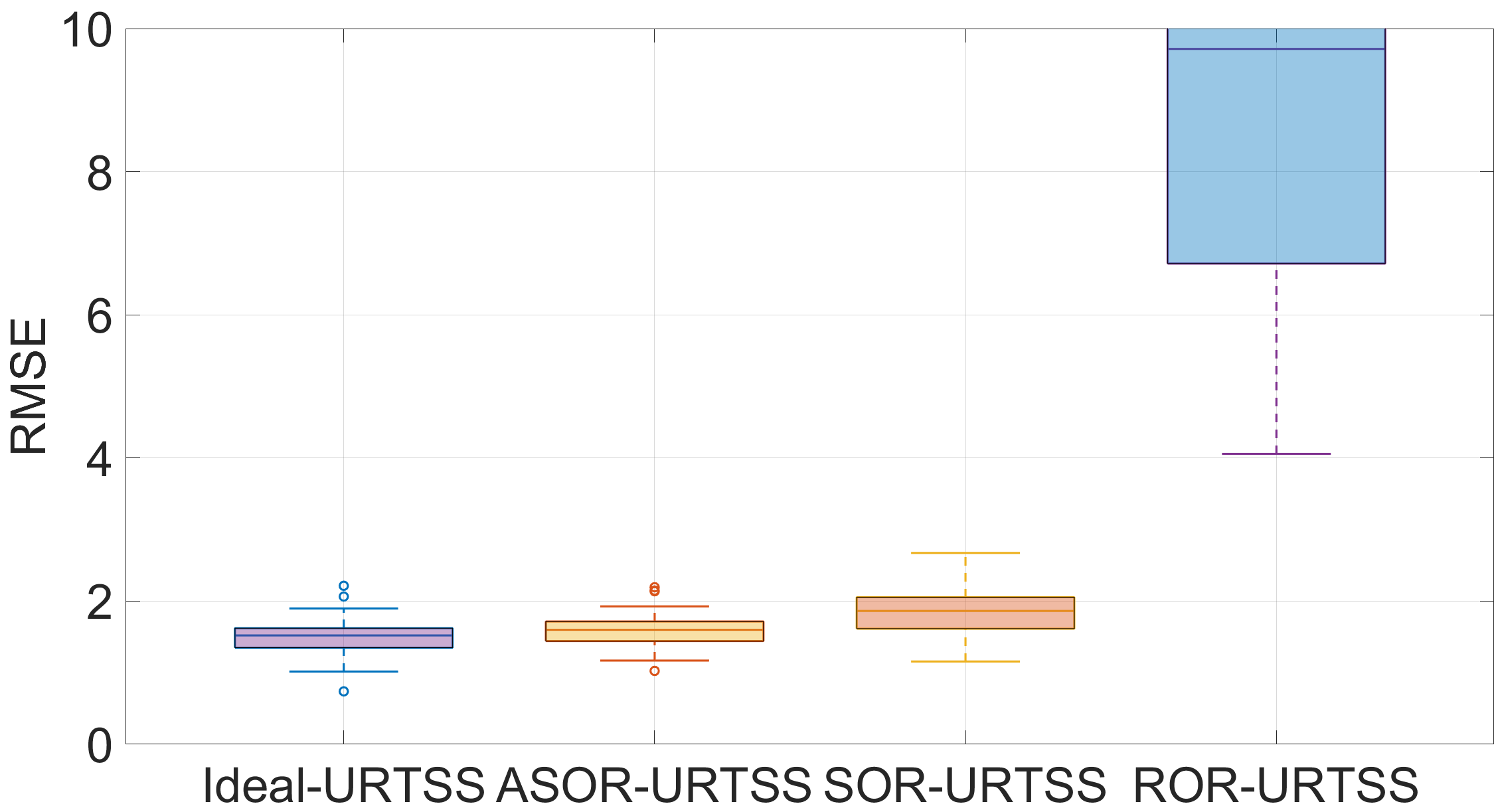}
        \subcaption{$m = 100$}\label{fig:b2}
    \end{minipage}\hfill
    \begin{minipage}[t]{0.24\textwidth}
        \centering
        \includegraphics[width=\textwidth]{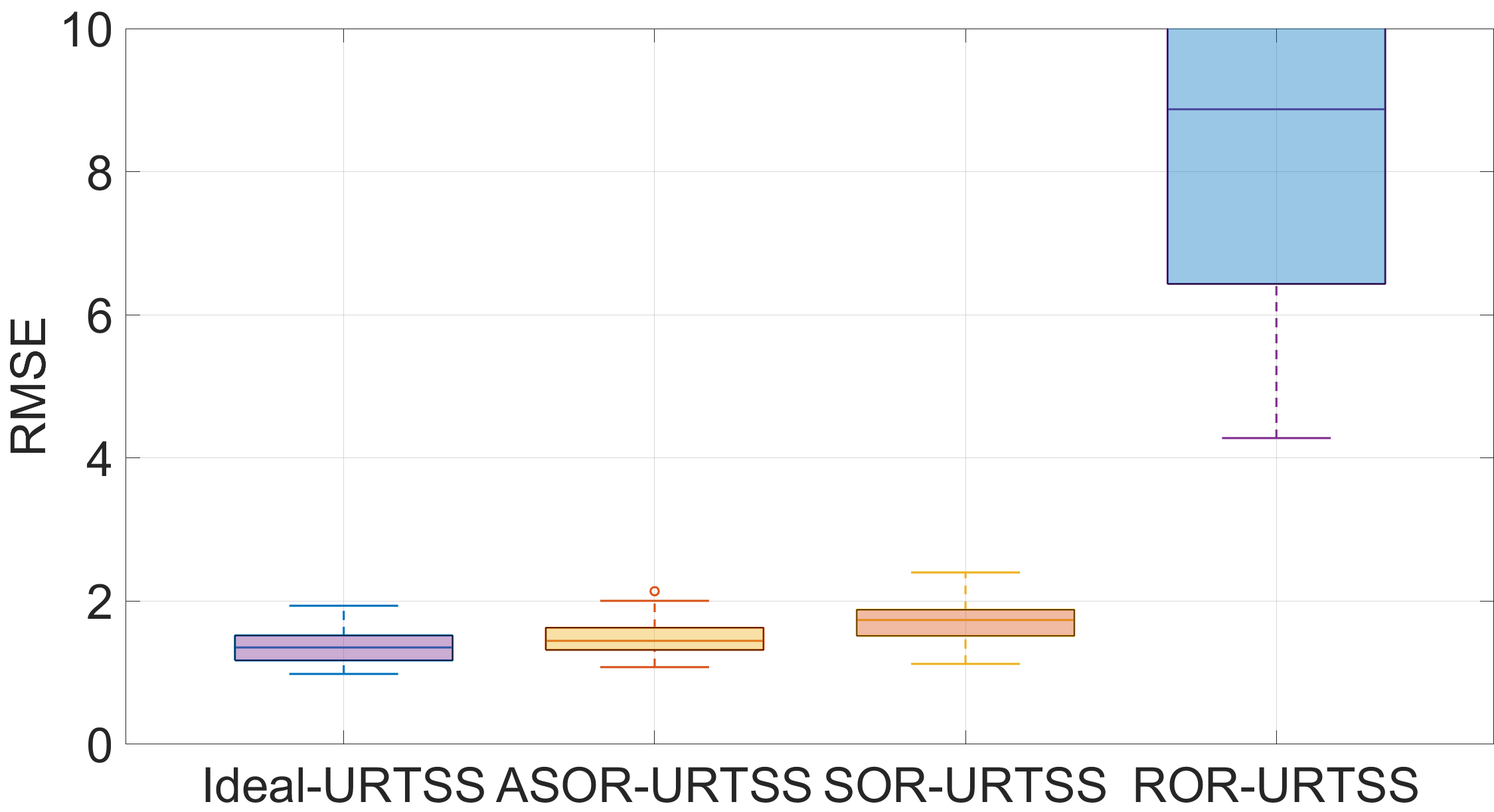}
        \subcaption{$m = 150$}\label{fig:c1}
    \end{minipage}\hfill
    \begin{minipage}[t]{0.24\textwidth}
        \centering
        \includegraphics[width=\textwidth]{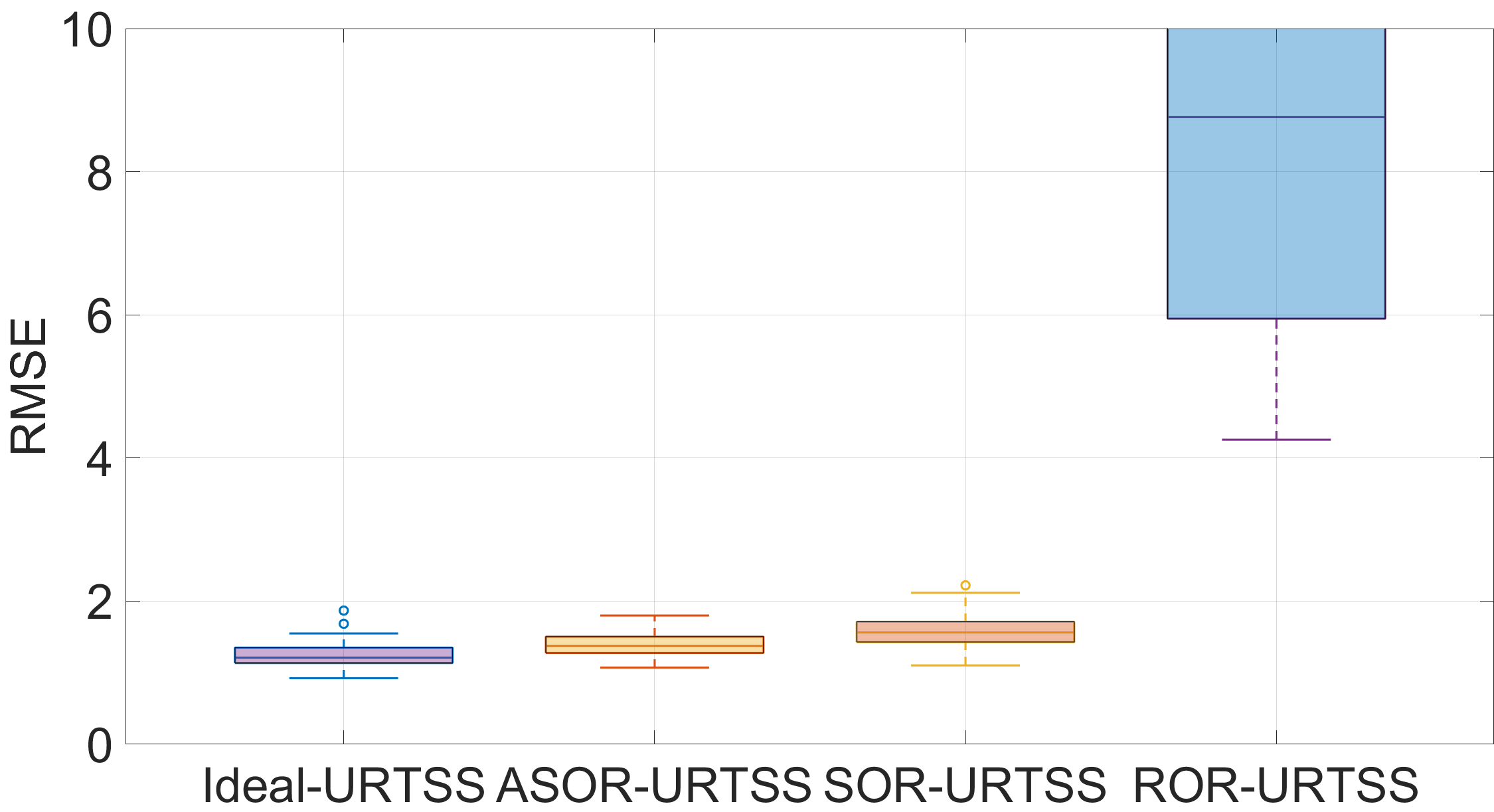}
        \subcaption{$m = 200$}\label{fig:d1}
    \end{minipage}
    
    \caption{Results for Simulation 2}
    \label{fig:s2}
\end{figure*}

This simulation demonstrates the gains in the performance for each algorithm by increasing the number of sensors i.e $m \in \{50, 100, 150, 200\}$. For this simulation we fix $\lambda = 0.4$. The results presented in Figure \ref{fig:s2} emphasize that our proposed method has the smallest RMSE for each value of $m$ compared to all other non-ideal smoothers.

\subsection{Simulation 3}
\begin{figure}[ht]
    \centering
    \includegraphics[width=0.4\textwidth]{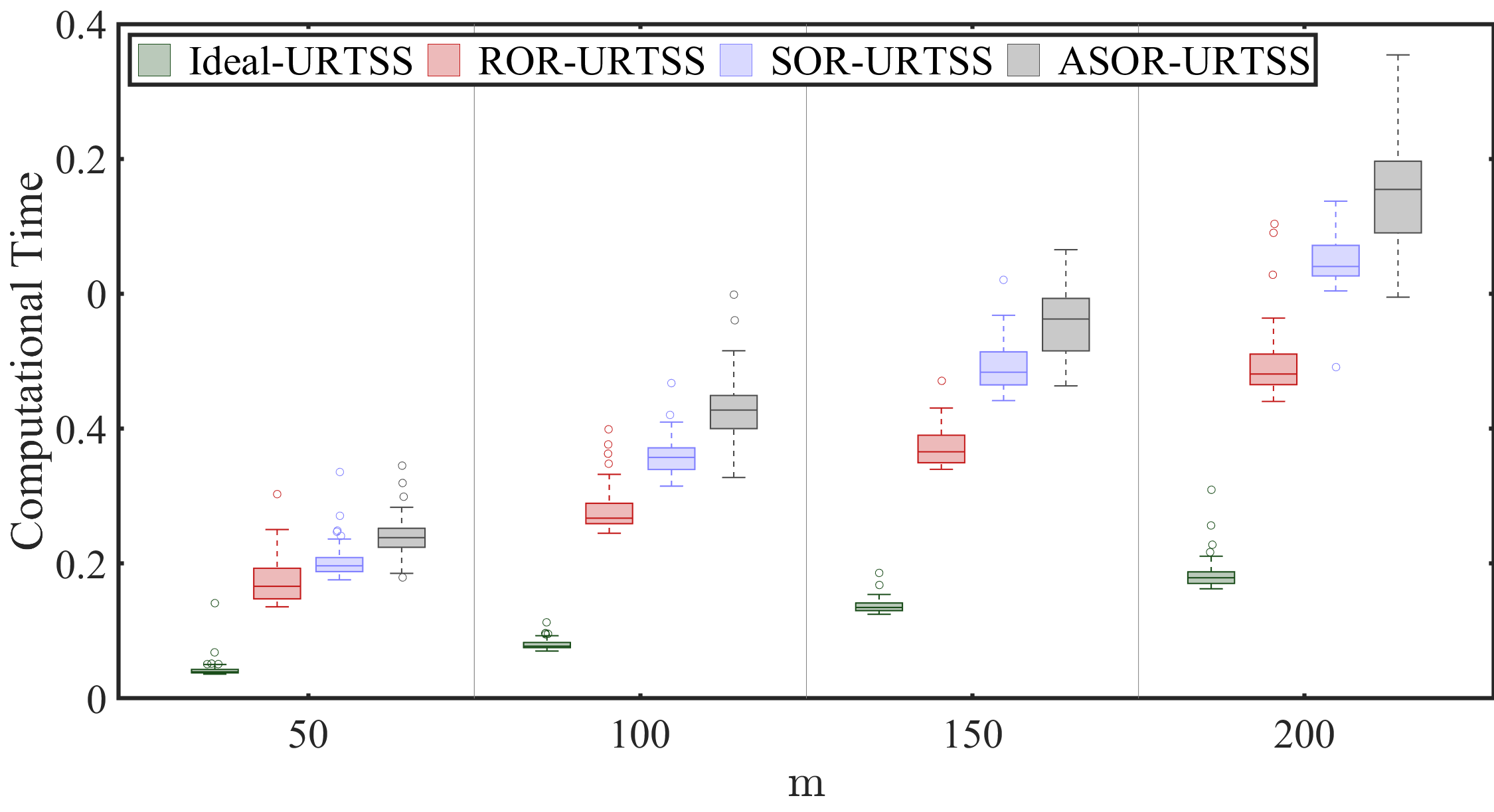} 
    \caption{Results for Simulation 3}
    \label{fig:s3}
\end{figure}

This simulation demonstrates that implementing our smoothers using the S-SPKF framework results in linear time complexity with respect to the number of sensors. The parameters for this simulation are identical to those in "Simulation 2", and the results are presented in Figure \ref{fig:s3}. While ASOR-URTSS exhibits higher computational time compared to SOR-URTSS, this is due to the additional requirement of estimating $\hat{b}_k$ at each time step, alongside the parameters common with SOR-URTSS.

\subsection{PIF simulation}
\begin{figure*}[ht]
    \centering
    \begin{minipage}[t]{0.24\textwidth}
        \centering
        \includegraphics[width=\textwidth]{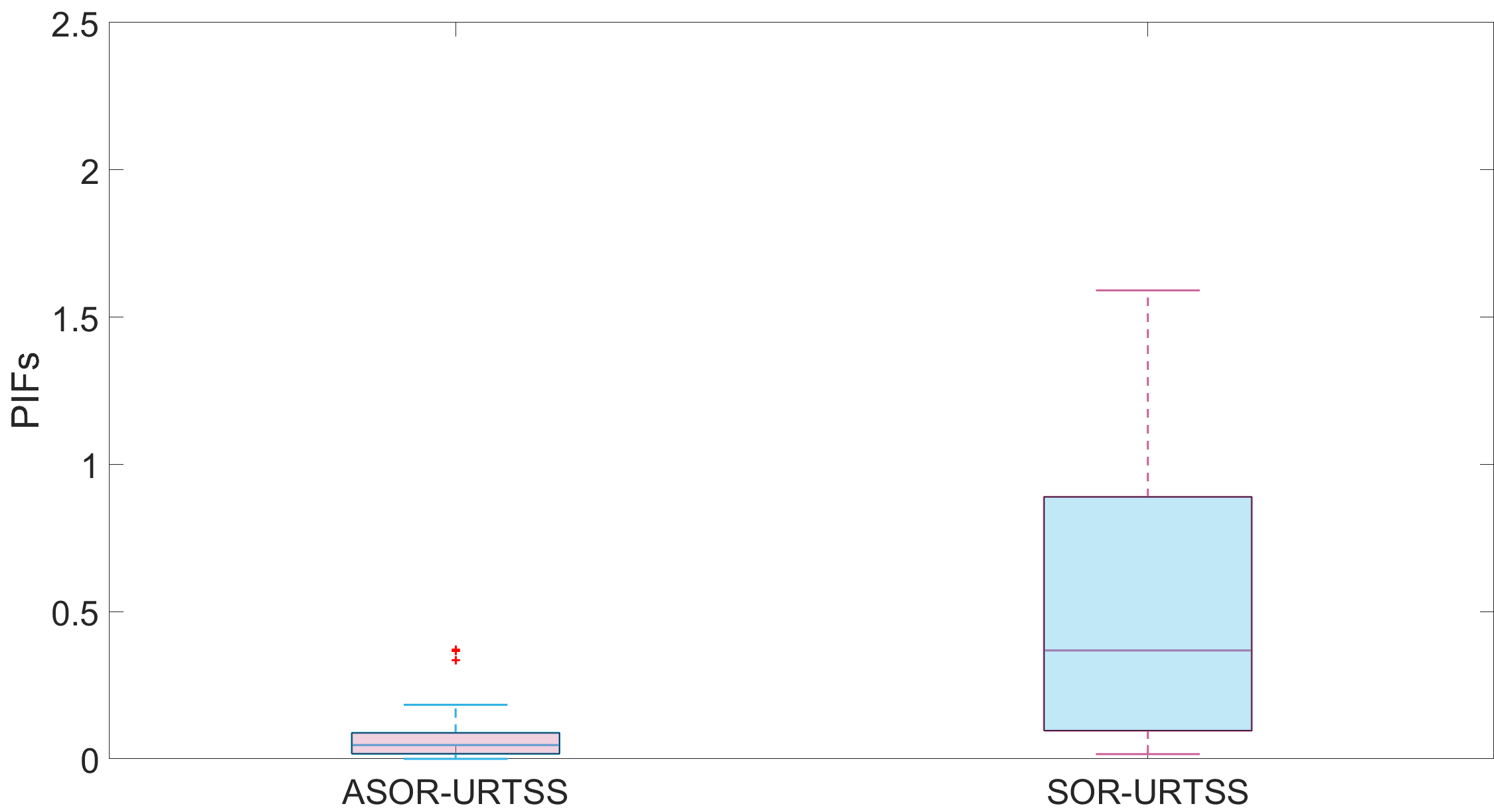}
        \subcaption{$\varsigma = \sqrt{10}$}\label{fig:a3}
    \end{minipage}\hfill
    \begin{minipage}[t]{0.24\textwidth}
        \centering
        \includegraphics[width=\textwidth]{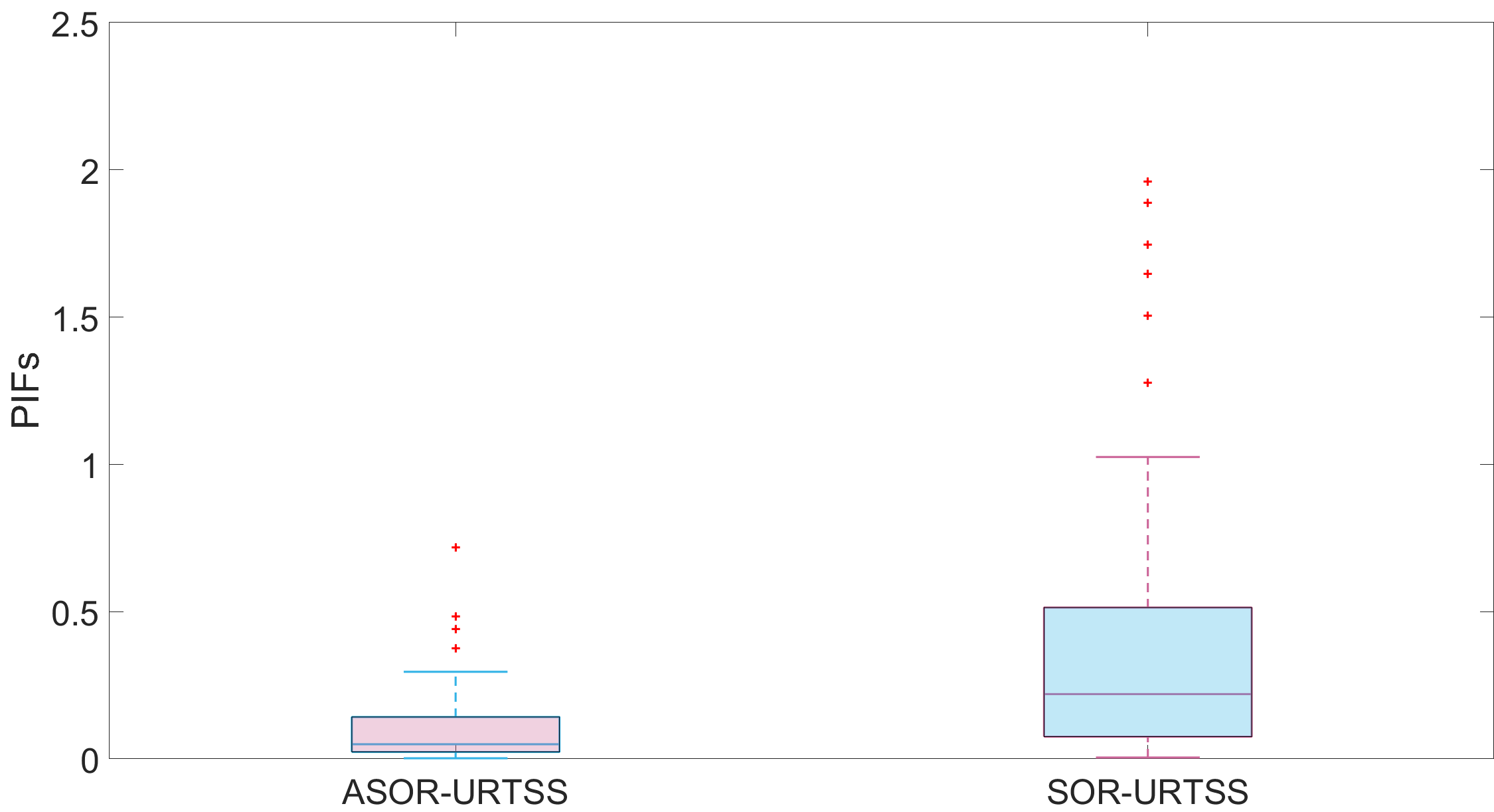}
        \subcaption{$\varsigma = \sqrt{100}$}\label{fig:b3}
    \end{minipage}\hfill
    \begin{minipage}[t]{0.24\textwidth}
        \centering
        \includegraphics[width=\textwidth]{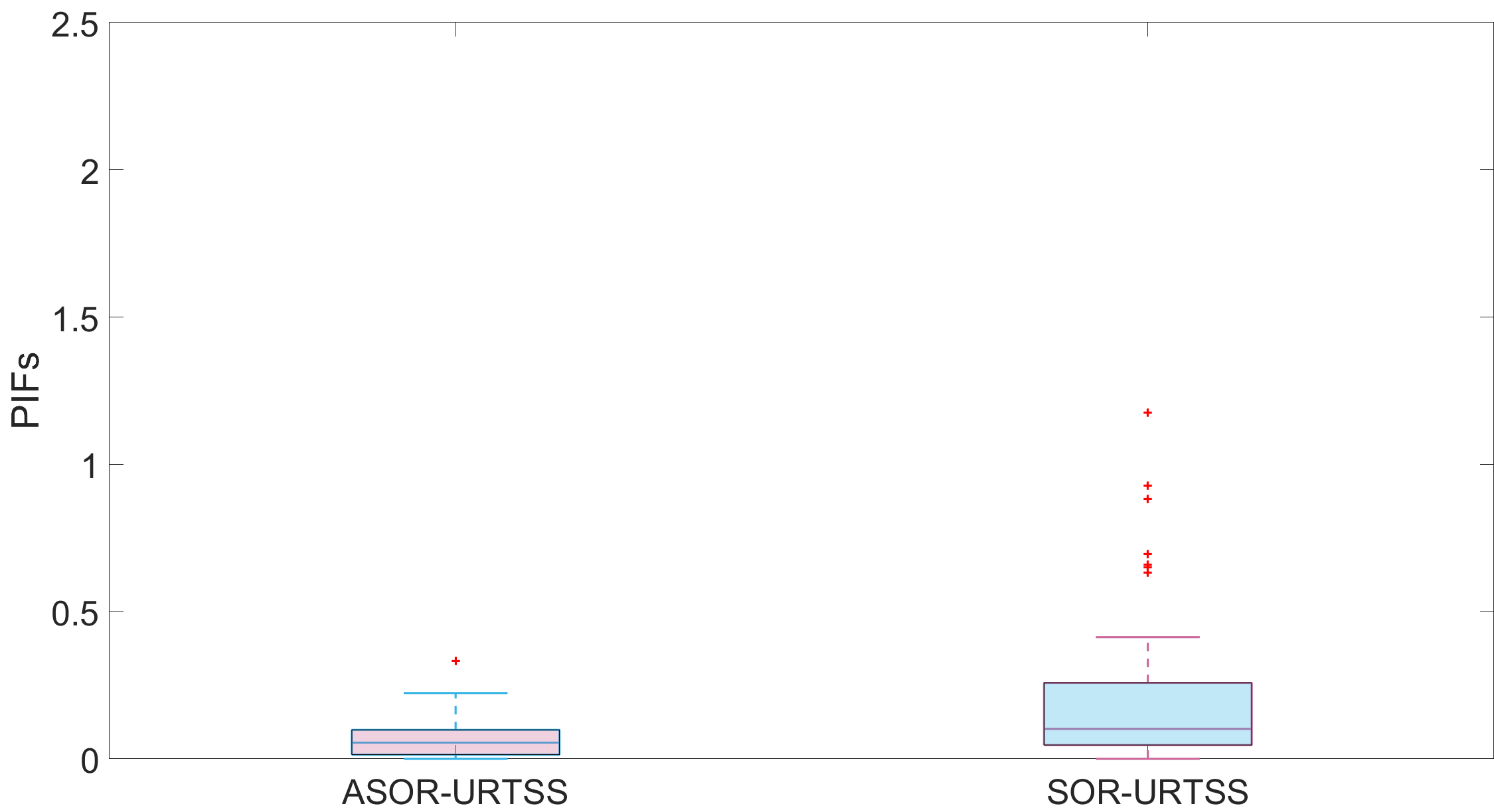}
        \subcaption{$\varsigma = \sqrt{1000}$}\label{fig:c2}
    \end{minipage}\hfill
    \begin{minipage}[t]{0.24\textwidth}
        \centering
        \includegraphics[width=\textwidth]{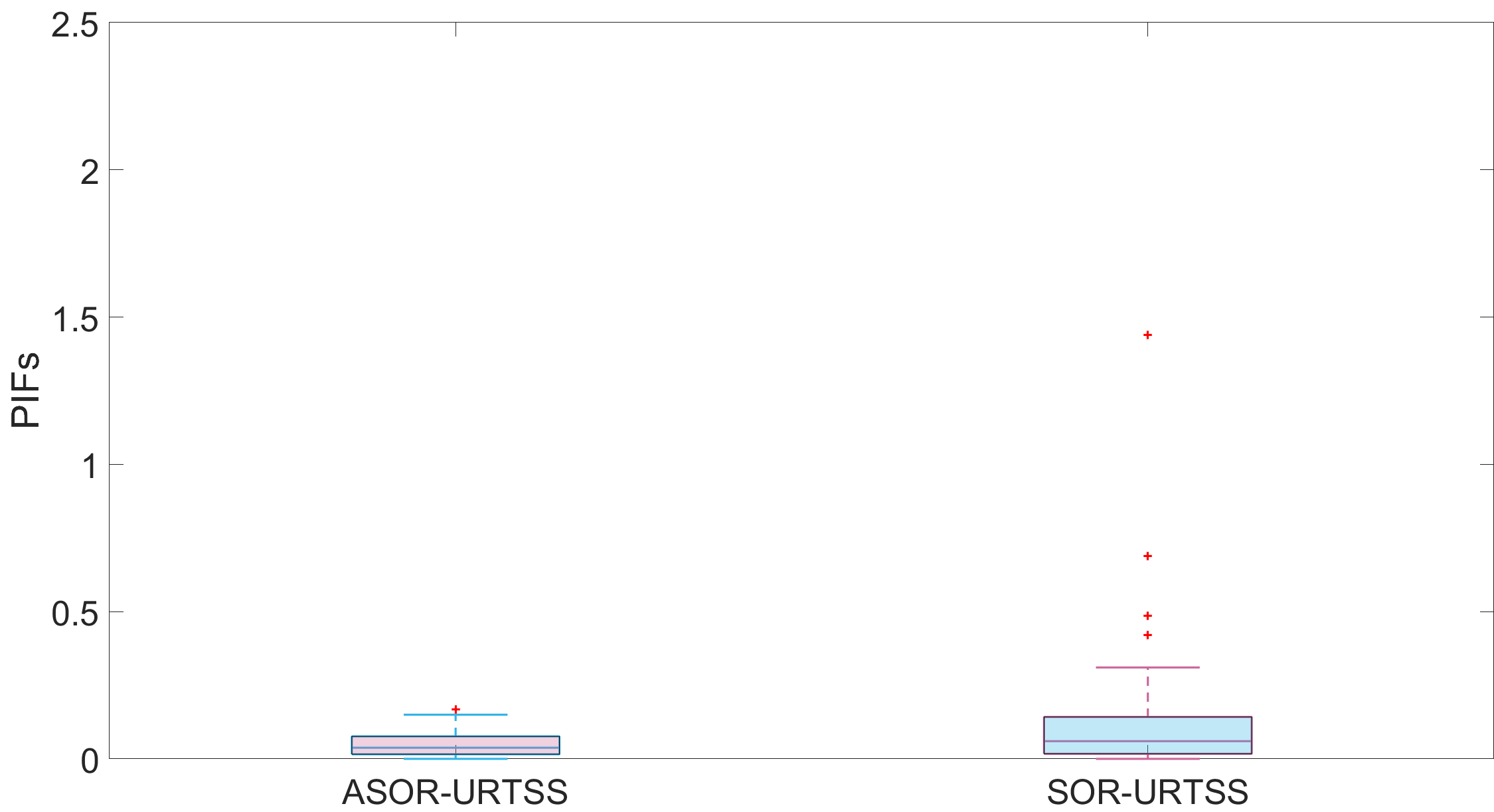}
        \subcaption{$\varsigma =\sqrt{10000}$}\label{fig:d2}
    \end{minipage}
    
    \caption{Results for PIF Simulation}
    \label{fig:pif}
\end{figure*}

We conducted simulations to compare the PIFs of the SOR-URTSS and ASOR-URTSS smoothers. The PIFs for both smoothers were estimated using \eqref{pif}. To evaluate the PIFs, a randomly generated outlier was introduced in all dimensions of the measurement vector at one random time step in each Monte Carlo run. The parameters of the smoothers were kept identical to those used in "Simulation 1", with the error variance, denoted by $\varsigma$, increased to simulate a larger discrepancy between $\mathbf{y}_k$ and $\mathbf{y}_k^c$. The smoothers were run for 50 Monte Carlo Runs for each value of $\varsigma$ and the results of these simulations are shown in Figure \ref{fig:pif}. The results suggest that the ASOR-URTSS has better robust characteristics compared to the SOR-URTSS across all the simulated scenarios.

\subsection{Initializing Weights}
To fully comply with the second criterion, \( \mathcal{I}_k^i \|\mathbf{y}_k\|_2 < \infty \), we cannot initialize our weights at 1. This is because the first forward pass lacks a robustness mechanism, and the algorithm relies on entering the first backward pass to activate the robustness mechanism. A large outlier during the initial forward pass can cause the algorithm to diverge, potentially resulting in \( \mathcal{I}_k^i \|\mathbf{y}_k\|_2 \rightarrow \infty \) for significant outliers in \( \mathbf{y}_k \).

\begin{figure*}[ht]
    \centering
    \begin{minipage}[t]{0.24\textwidth}
        \centering
        \includegraphics[width=\textwidth]{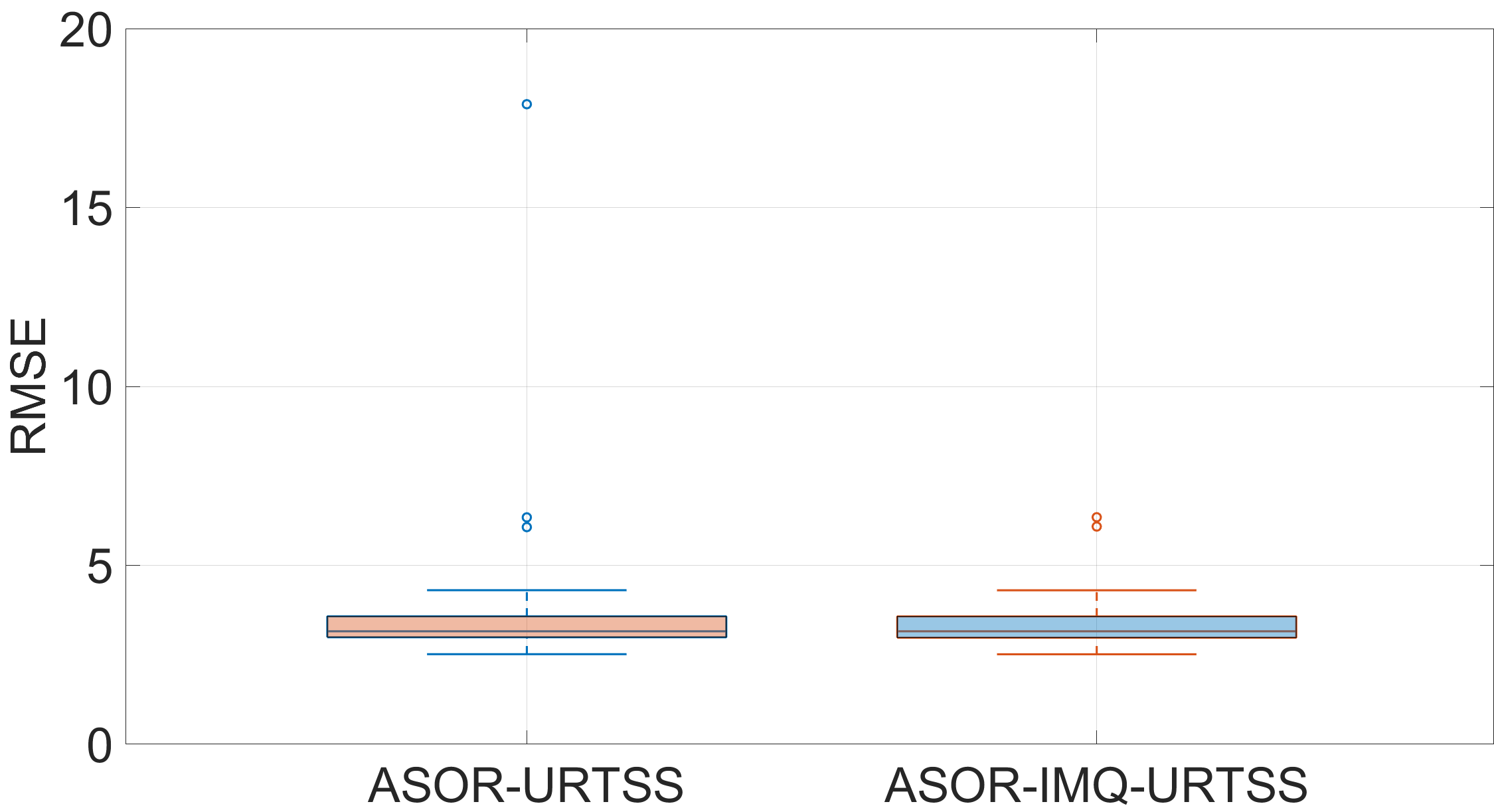}
        \subcaption{$\lambda = 0.2$}\label{fig:a4}
    \end{minipage}\hfill
    \begin{minipage}[t]{0.24\textwidth}
        \centering
        \includegraphics[width=\textwidth]{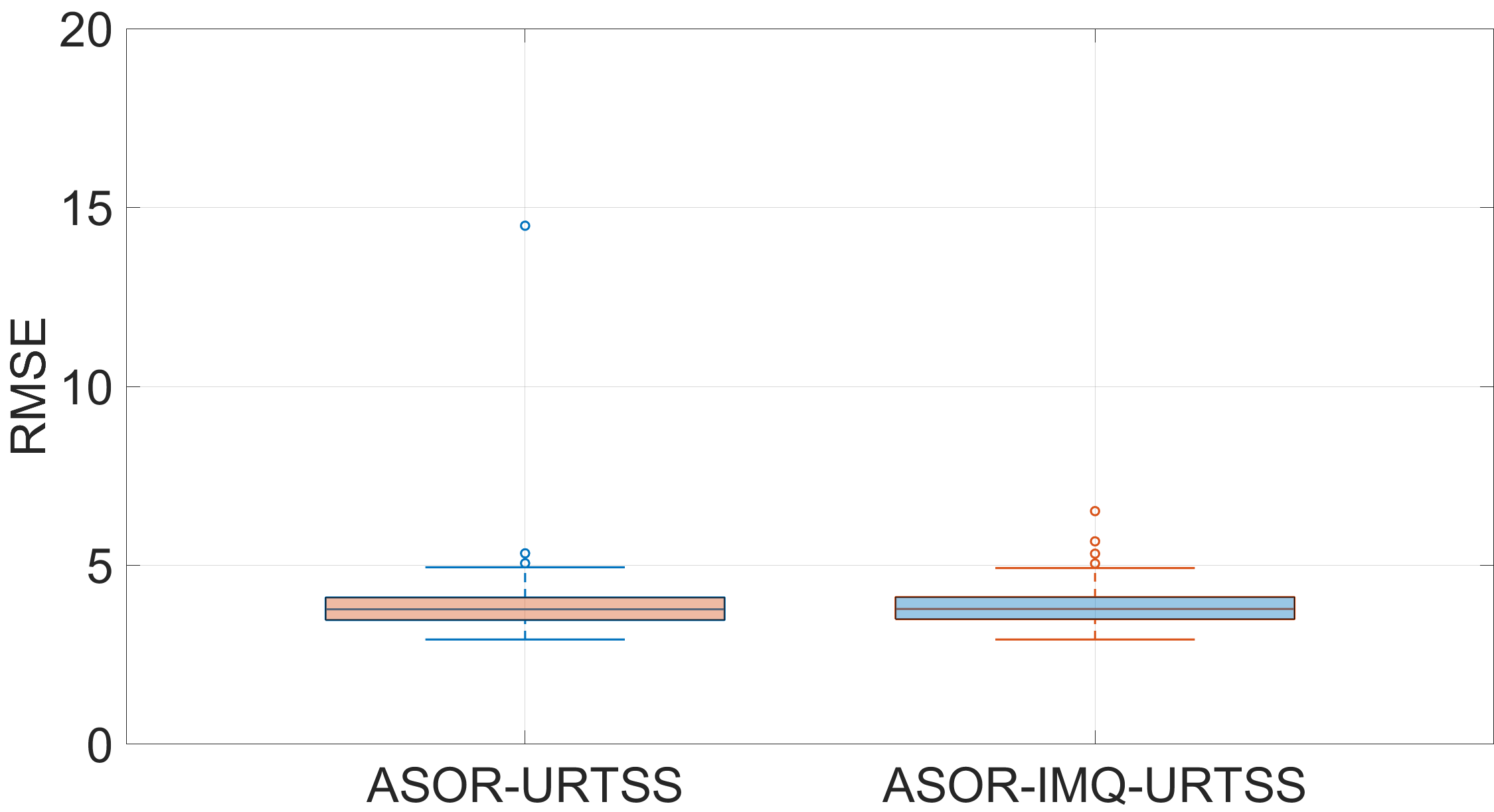}
        \subcaption{$\lambda = 0.4$}\label{fig:b4}
    \end{minipage}\hfill
    \begin{minipage}[t]{0.24\textwidth}
        \centering
        \includegraphics[width=\textwidth]{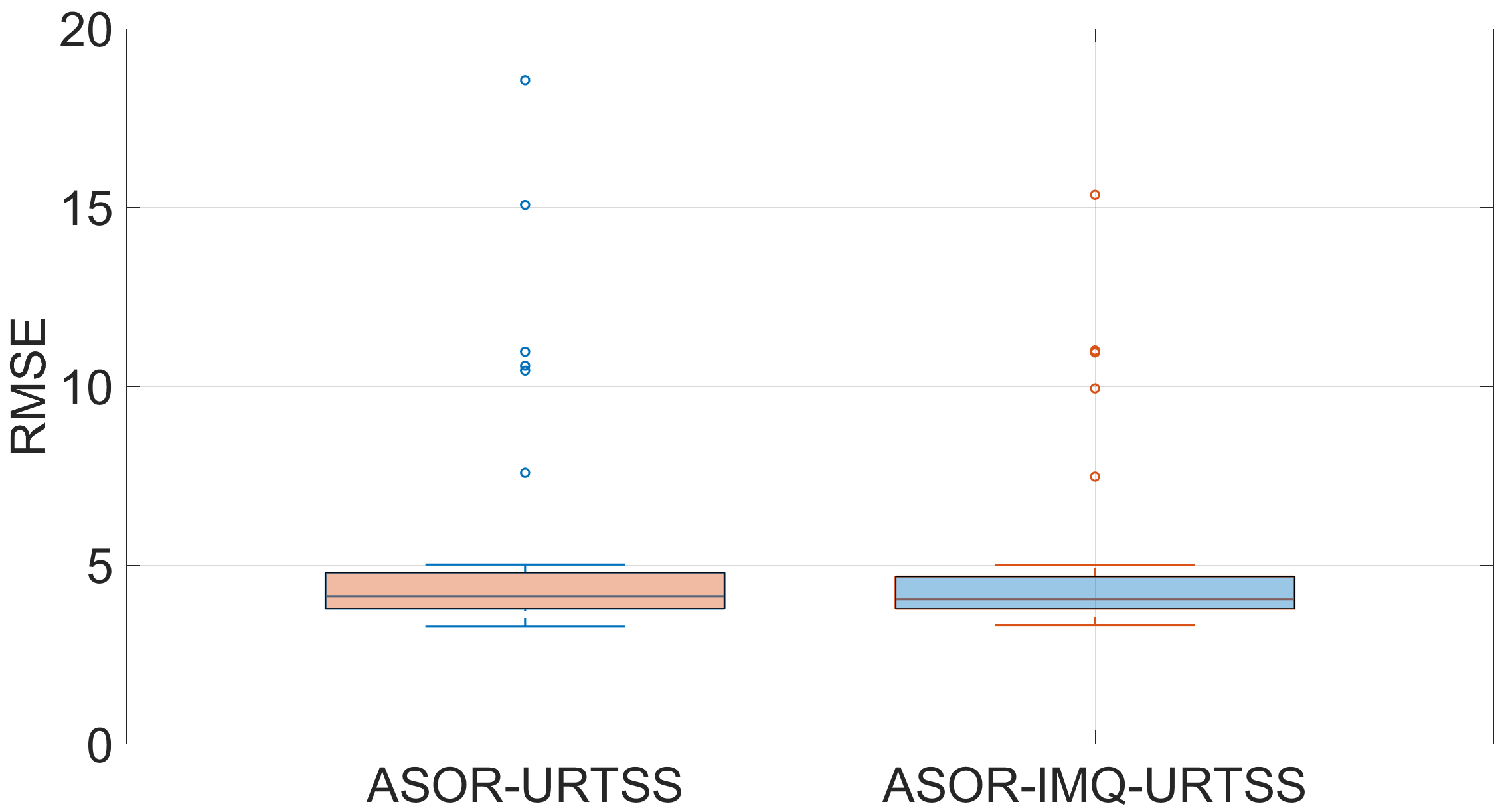}
        \subcaption{$\lambda = 0.5$}\label{fig:c3}
    \end{minipage}\hfill
    \begin{minipage}[t]{0.24\textwidth}
        \centering
        \includegraphics[width=\textwidth]{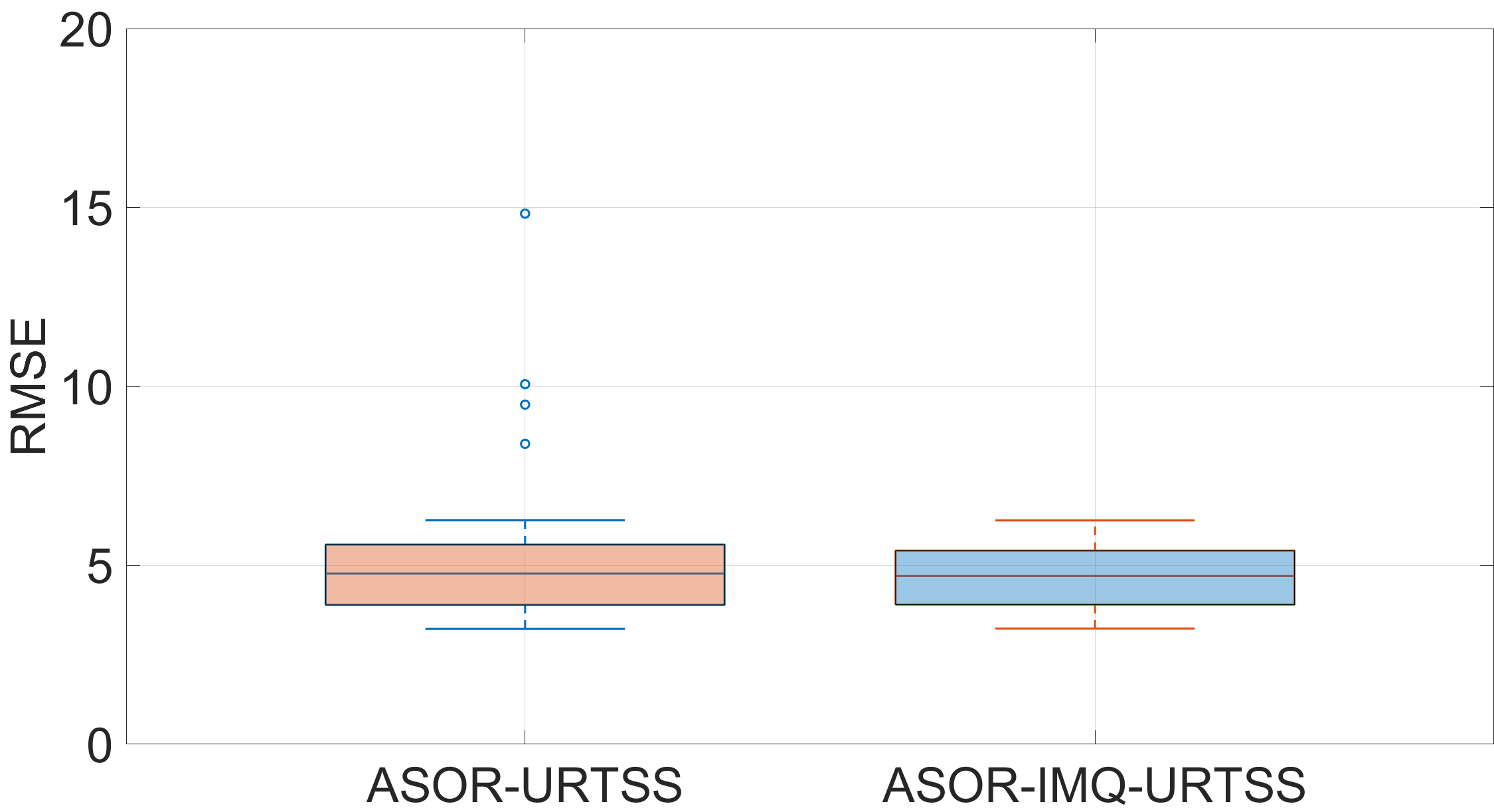}
        \subcaption{$\lambda = 0.6$}\label{fig:d3}
    \end{minipage}
    
    \caption{Comparing ASOR-URTSS and ASOR-IMQ-URTSS}
    \label{fig:imq}
\end{figure*}

While we did not encounter many such divergences in our tests and simulations, it is essential to use initial weights that maintain the PIF robustness criterion, even during the first forward pass. To address this, we adopted the inverse multi-quadratic (IMQ) \cite{altamirano2023robust} weighting function to derive our weights during the first forward pass . This ensures robustness from the start of the process.

\begin{align}
    &W_k=\left(1+\frac{||\mathbf{y}_k-\mathbf{h}(\mathbf{x}_k)||_2^2}{c^2}\right)^{-1/2}
\end{align}
In this context, \( c \) represents the soft threshold. During the first forward pass, we can use the weights \( W_k \) instead of \( \mathbf{\mathcal{I}}_k \), since \( W_k \|\mathbf{y}_k\|_2 < \infty \), even in the presence of large outliers in \( \mathbf{y}_k \). To demonstrate that using \( W_k \) does not affect the overall performance and convergence of the ASOR-URTSS, we present the results of a simulation with parameters similar to those used in "Simulation 1" with \( T = 500 \) and \( c = 5 \) in Figure \ref{fig:imq}. The experiment was repeated for 50 Monte Carlo runs for each value of \( \lambda \). The version of the smoother using \( W_k \) for the first forward pass is referred to as ASOR-IMQ-URTSS.

As can be observed, the performance of both smoothers is comparable. However, in exceptional cases, ASOR-IMQ-URTSS exhibits slightly lower RMSE values. This improvement may be attributed to the prevention of smoother divergence during the initial forward pass.

\section{Results on Real Data}
We evaluate our algorithm on real data from \cite{chughtai2022outlier}, consisting of UWB range measurements collected using the Qorvo MDEK1001 Development Kit. The firmware controls the UWB transceivers to form an anchor network and perform two-way ranging with tag nodes, allowing each tag to calculate its relative location. Range data is logged at 5Hz from up to 4 nearest anchors and averaged for final readings, with the tag following a known path. Outliers arise from missing measurements (up to 4 of 11 anchors provide data at a time) and bias due to non-line-of-sight (NLoS) obstructions.
\\
\subsection{System Model}

We consider a random walk as the state mobility dynamic model, commonly used for inference in various applications, including mobile nodes in wireless sensor networks \cite{article2}. The 2D position of the target is our quantity of interest, and the state vector is defined as:
$\mathbf{x}_k = [a_k, b_k]^\top$
The state evolves with a simple identity function, \(\mathbf{f}(.) = \mathbf{I}\), and the nominal measurement model is given by:
\[
\mathbf{h}(\mathbf{x}_k) = \sqrt{(a_k - a^{\rho_j})^2 + (b_k - b^{\rho_j})^2}
\]
where a term for the z-axis of the tag and anchor locations is implicitly included. The process noise covariance \(\mathbf{Q}_{k}\) and measurement noise covariance \(\mathbf{R}_k\) are diagonal matrices with diagonal entries set to 0.1.

For each case, the initial state \(\mathbf{x}_0\) is sampled from the distribution:
$
\mathbf{x}_0 \sim \mathcal{N} \left([x_1, x_2]^\top, \vartheta \mathbf{Q}_k \right)
$
where \(\vartheta = 10\), \((x_1, x_2)\) represent the actual initial position from the dataset, and the initial covariance is \(\mathbf{P}_0 = \vartheta \mathbf{Q}_k\).

\begin{table}[ht]
    \centering
    \caption{RMSE Results for Different Scenarios}
    \begin{tabular}{|c|c|c|c|}
        \hline
        \textbf{Scenario} & \textbf{SOR-URTSS} & \textbf{ASOR-URTSS} & \textbf{ROR-URTSS} \\ \hline
        \textbf{1}        & 0.3657             & 0.3613              & 6.5182             \\ \hline
        \textbf{2}        & 0.3778             & 0.3793              & 3.9911             \\ \hline
        \textbf{3}        & 0.4232             & 0.4247              & 3.4216             \\ \hline
    \end{tabular}
    \label{tab:rmse_results}
\end{table} 
\subsection{Discussion on Results}
\begin{figure*}[ht]
    \centering
    \begin{minipage}[t]{0.32\textwidth}
        \centering
        \includegraphics[width=\textwidth]{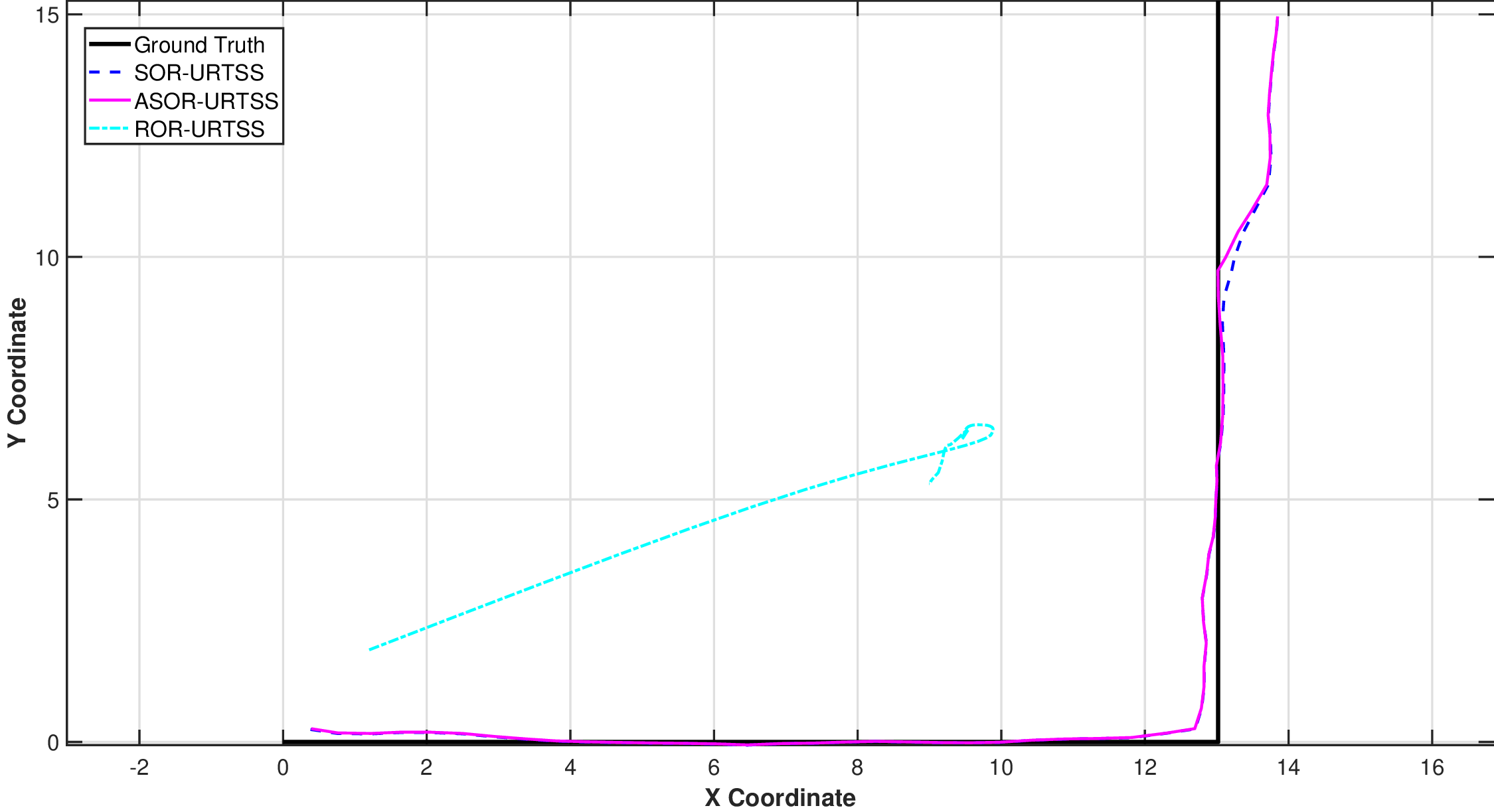}
        \subcaption{Scenario 1}\label{fig:a5}
    \end{minipage}\hfill
    \begin{minipage}[t]{0.32\textwidth}
        \centering
        \includegraphics[width=\textwidth]{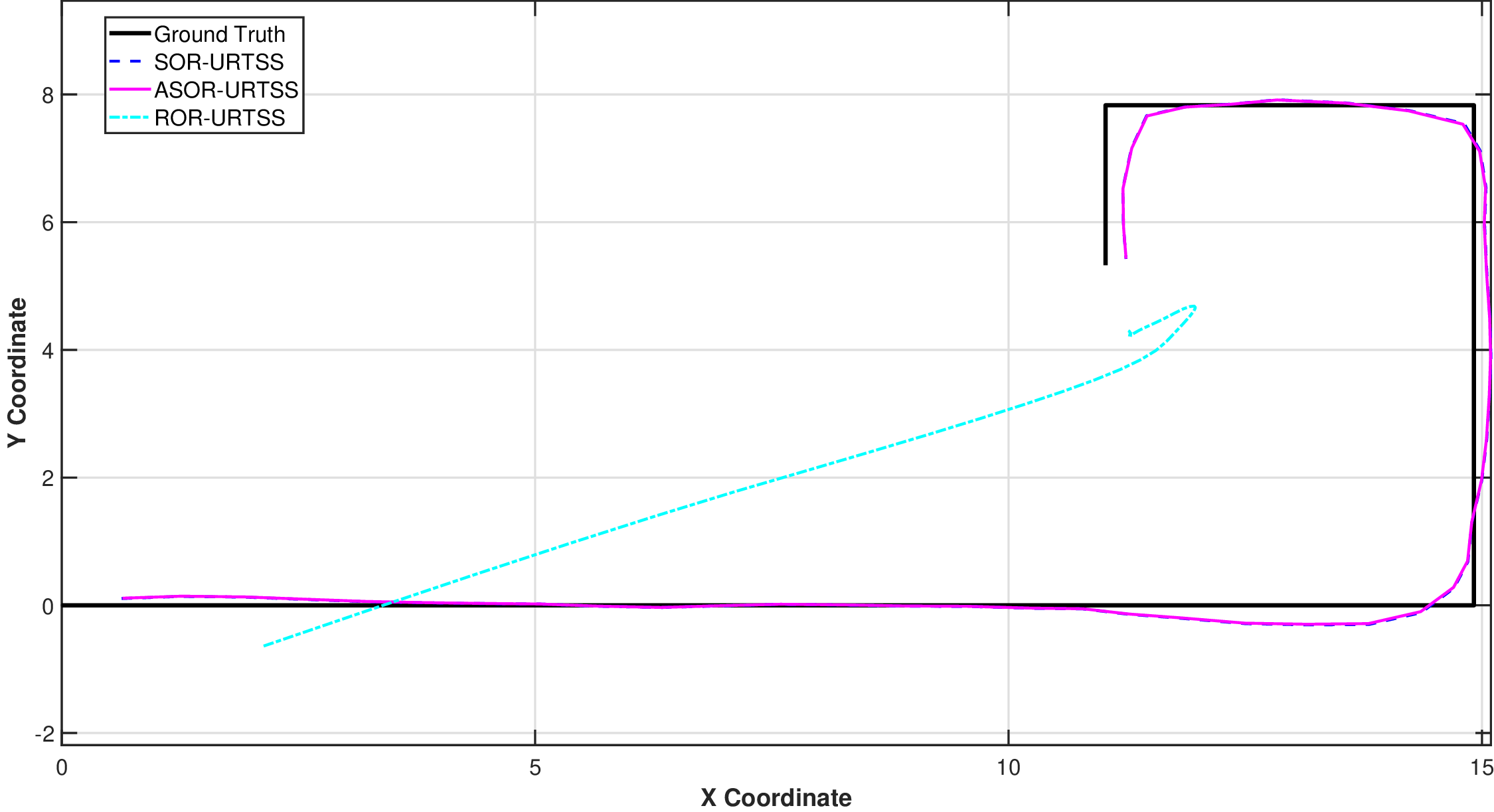}
        \subcaption{Scenario 2}\label{fig:b5}
    \end{minipage}\hfill
    \begin{minipage}[t]{0.32\textwidth}
        \centering
        \includegraphics[width=\textwidth]{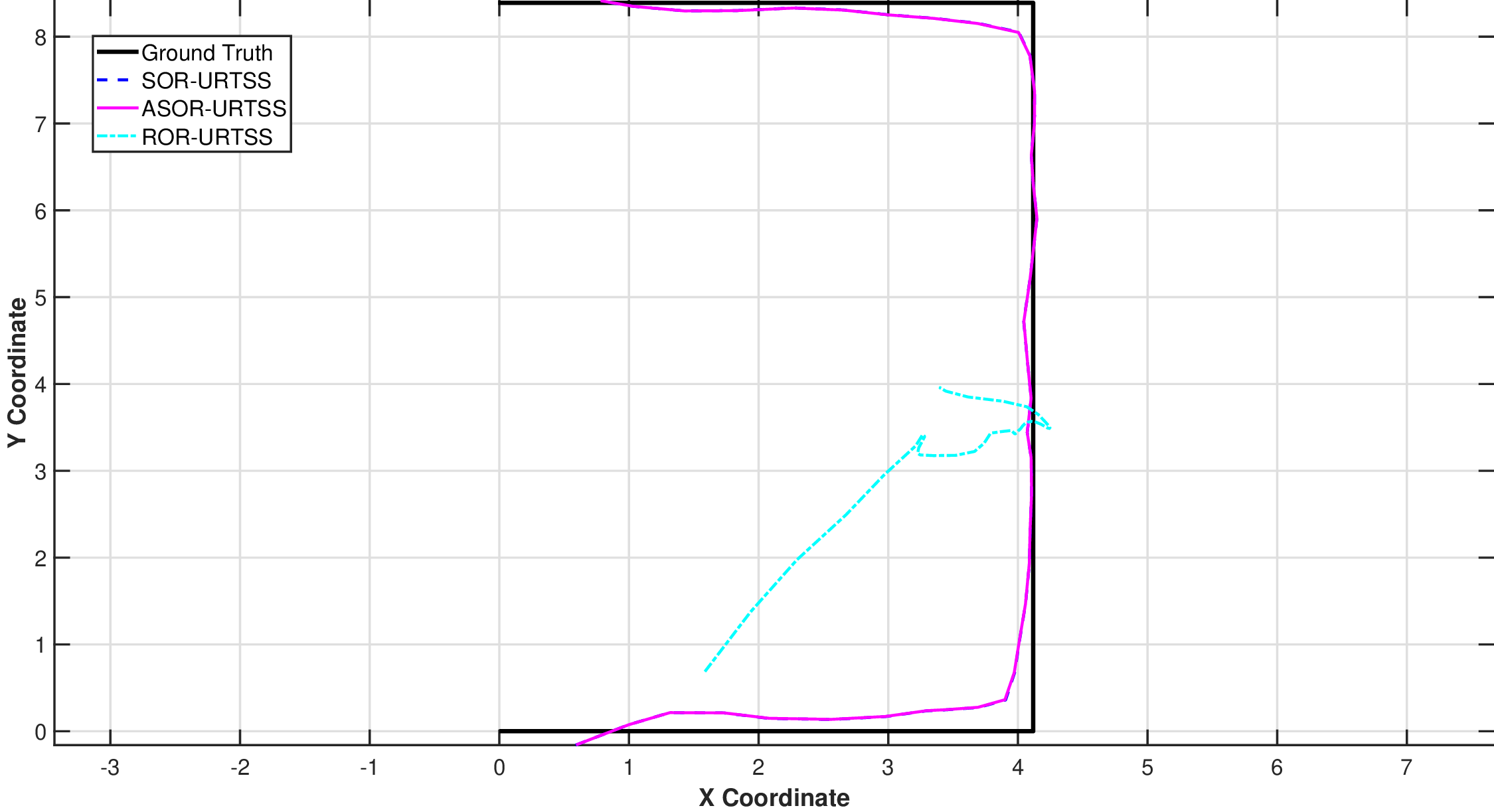}
        \subcaption{Scenario 3}\label{fig:c4}
    \end{minipage}
    
    \caption{Tracking Plots}
    \label{fig:tp}
\end{figure*}

We present the results of our proposed smoother, along with comparative techniques, in Figure \ref{fig:tp}, with the corresponding RMSE values summarized in Table \ref{tab:rmse_results}. The results clearly indicate that the ASOR-URTSS closely follows the performance of the SOR-URTSS in terms of RMSE. We understand this similarity in error performance to be a result of an inherent lack of changing error characteristics in the data.

To further evaluate the robustness of our proposed method, we conducted additional tests on real-world data, comparing ASOR-URTSS against SOR-URTSS. These tests demonstrate that ASOR-URTSS is notably more resilient to variations in initial uncertainty. Specifically, we observed that increasing the initial uncertainty, denoted by $\vartheta$, can lead to frequent divergence in the SOR-URTSS, whereas ASOR-URTSS remains significantly more stable.

\begin{figure*}[ht]
    \centering
    \begin{minipage}[t]{0.32\textwidth}
        \centering
        \includegraphics[width=\textwidth]{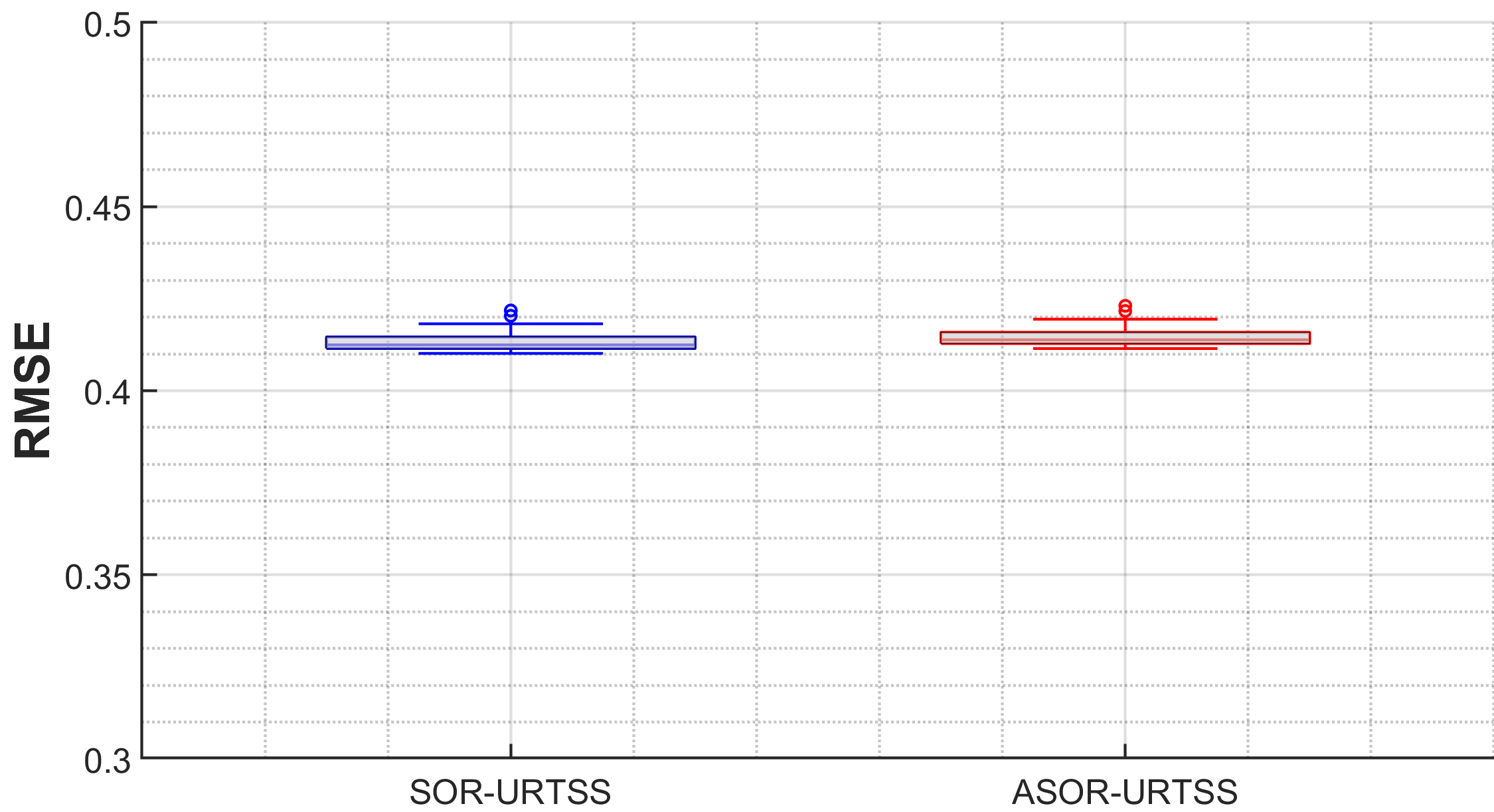}
        \subcaption{$\vartheta = 1$}\label{fig:a6}
    \end{minipage}\hfill
    \begin{minipage}[t]{0.32\textwidth}
        \centering
        \includegraphics[width=\textwidth]{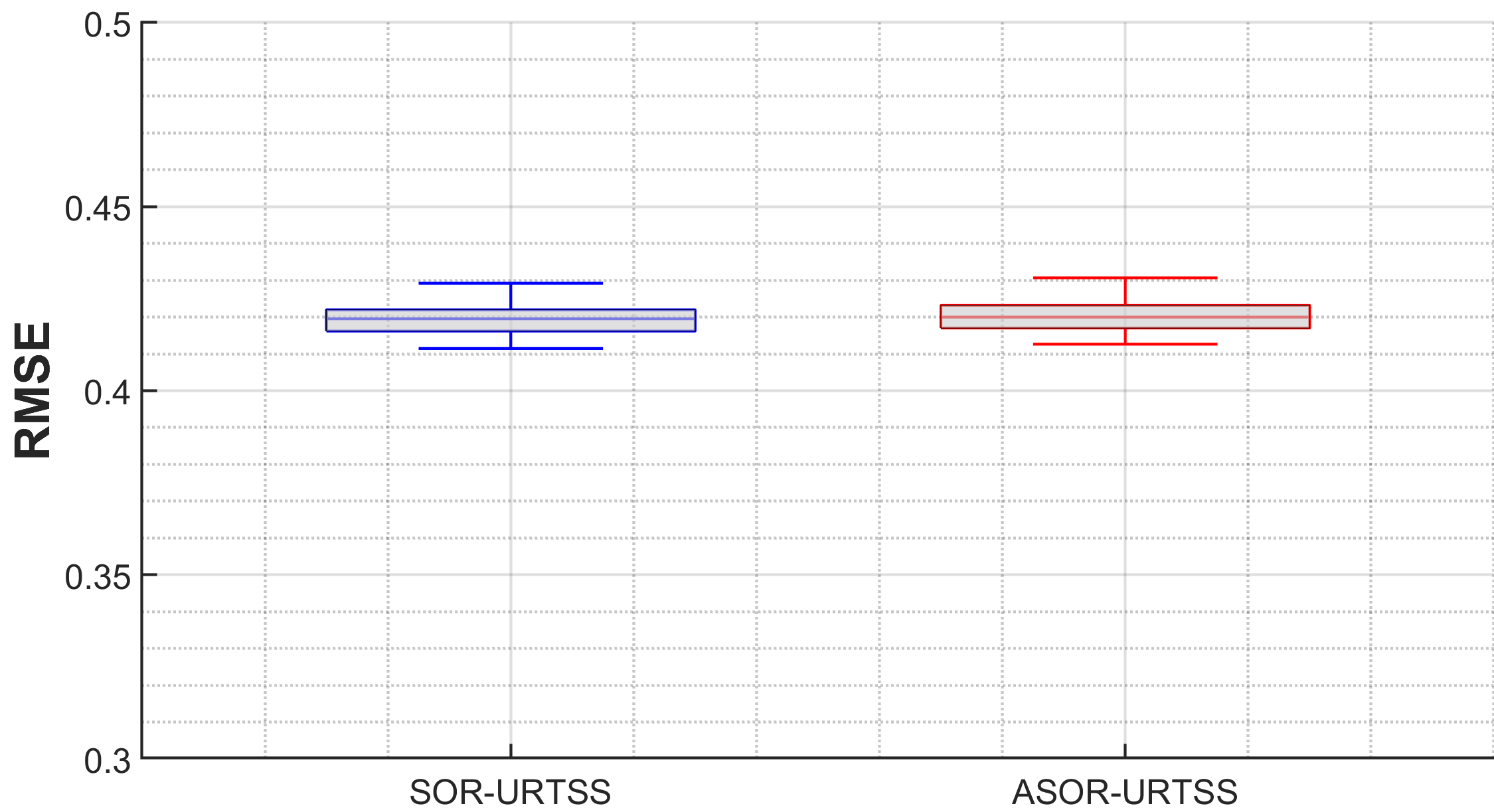}
        \subcaption{$\vartheta = 10$}\label{fig:b6}
    \end{minipage}\hfill
    \begin{minipage}[t]{0.32\textwidth}
        \centering
        \includegraphics[width=\textwidth]{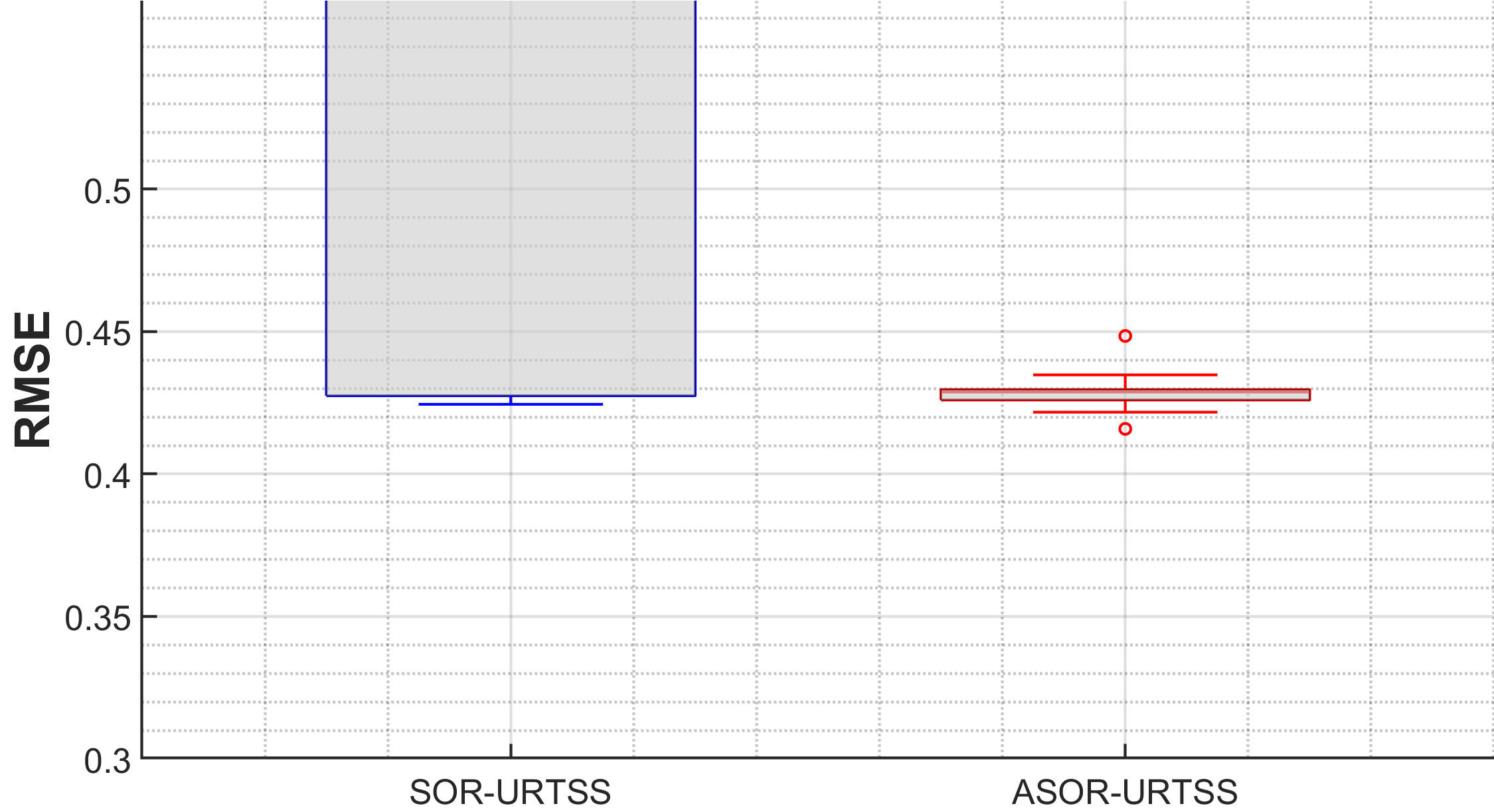}
        \subcaption{$\vartheta = 100$}\label{fig:c5}
    \end{minipage}
    
    \caption{Tracking Plots for Increasing Initial Uncertainty}
    \label{fig:tim}
\end{figure*}

To illustrate this point, we present the results of increasing $\vartheta \in \{1, 10, 100\}$ in "Scenario 3" over 50 Monte Carlo runs, as shown in Figure \ref{fig:tim}. While the RMSE of ASOR-URTSS remains very close to that of SOR-URTSS, the stability of ASOR-URTSS under increasing uncertainty makes it a more reliable option for scenarios involving high initial uncertainty.

\section{Conclusion}

In this article, the TR problem is treated from an offline Gaussian smoothing perspective. The RTS framework for smoothing with a forward and backward pass is adopted. A VB based selective outlier rejecting algorithm is proposed which utilizes a hierarchical Bayesian model to detect and mitigate outliers in the measurement vector. The model employs a Gamma distribution to dynamically learn the characteristics of outliers in the data allowing for a more scrupulous suppression of outliers during each-time step. The article further presents a robustness criterion for smoothers based on PIFs and argues that the proposed smoother is robust based on this criterion. Results on simulation are presented to showcase performance gains in comparison to other VB based robust smoothing methods and also establish the linear time complexity in terms of the number of sensors of the proposed method. Results on real data from UWBs is also presented to demonstrate encouraging  applicability in real scenarios. Further simulations targeting the proposed robustness criterion are presented to illustrate superior robustness characteristics compared to another selective outlier rejecting smoother.

\bibliographystyle{ieeetr} 

\end{document}